\documentclass[useAMS,usenatbib]{mn2e}

\usepackage{graphicx}

\title[Models of Class 0 Outflow Sources]
{Unified Models of Molecular Emission from Class 0 Protostellar 
Outflow Sources}

\author[J.M.C. Rawlings et al.]
{J.M.C.~Rawlings,$^1$\thanks{E--mail: jcr@star.ucl.ac.uk}
M.P.~Redman$^2$ and P.B.~Carolan$^2$ \\
$^1$University College London, Department of Physics and Astronomy,  
Gower Street, London WC1E 6BT, United Kingdom\\
$^2$Centre for Astronomy, School of Physics, National University of Ireland
Galway, Galway, Ireland}

\begin{document}

\maketitle

\begin{abstract}

Low mass star-forming regions are more complex than the simple spherically
symmetric approximation that is often assumed. We apply a more realistic 
infall/outflow physical model to molecular/continuum observations of three
late Class 0 protostellar sources with the aims of (a) proving the 
applicability of a single physical model for all three sources, and (b) 
deriving physical parameters for the molecular gas component in each of the
sources.

We have observed several molecular species in multiple 
rotational transitions. The observed line profiles were modelled
in the context of a dynamical model which incorporates infall and 
bipolar outflows, using a three dimensional radiative transfer code. 
This results in constraints on the physical parameters and chemical
abundances in each source.

Self-consistent fits to each source are obtained. We constrain the 
characteristics of the molecular gas in the envelopes as well as in the
molecular outflows. We find that the molecular gas abundances in the 
infalling envelope are reduced, presumably due to freeze-out, whilst 
the abundances in the molecular outflows are enhanced, presumably due to
dynamical activity. 
Despite the fact that the line profiles show significant source-to-source
variation, which primarily derives from variations in the outflow viewing 
angle, the physical parameters of the gas are found to be similar
in each core. 

\end{abstract}


\section{Introduction}

The astrophysical mechanism of gravitational infall coupled with rotation 
and axisymmetric outflows operates on a vast range of scales, ranging from
black holes and jets in active galactic nuclei (such as M87), through 
to high and low mass star formation \citep{livio04} and 
even brown dwarfs \citep{whelan.et.al07}.

Searches for the signatures of the early stages of star formation
have tended to focus on infall indicators in Class 0 low mass protostellar 
objects. These possess a luminous protostellar core, but - by definition -
most of the mass is contained in a surrounding envelope. In most models the 
fraction of the mass in the envelope that is infalling increases with time
\citep[e.g.][]{shu.et.al87}. In the early stages the gas is approximately 
isothermal, and the envelope can be approximated by a spherically 
symmetric hydrostatically supported cloud. Emission from the envelope 
includes regions where the gas is infalling and and where it is
static.

In these circumstances, the classic observational signature of infalling 
optically thick molecular gas is a double peaked asymmetric line emission 
profile \citep{zhou.et.al93}. Molecular 
emission lines originating from quiescent
gas residing in the outer parts of the envelope have linewidths consistent
with thermal contributions alone as opposed to thermal and
turbulent line broadening of dynamic gas \citep{keto.et.al04}.

However, 
collapse seems to almost always lead promptly to disk, jet and (bipolar)
outflow formation. 
These outflows arise because molecular gas in the
envelope is entrained by protostellar jets. The latter are hot, high velocity ($\sim$ 200 km s$^{-1}$) jets of mostly
atomic hydrogen that carry away excess angular momentum from the
core. 
It is difficult to isolate cases of clouds on the verge of collapse
and in such cases local turbulent or global oscillations can mask the weak 
initial collapse motions \citep{lada.et.al03,redman.et.al06,gao&lou10}.
Once collapse is clearly present, line profile shapes become complex and 
ambiguous. The classic blue-asymmetric line profile shape that can indicate
collapse can also be generated by other large scale dynamical effects such as 
rotation \citep{redman.et.al04a} or
outflows \citep{rawlings.et.al04}. 
Moreover, whilst dynamic processes alter spectral line profiles they are
also affected by chemical processes which change the molecular
abundance, thereby altering the line intensity. These chemical
processes include gas-phase reactions, the freeze-out of molecular gas onto 
dust grains \citep{schoier.et.al02} in the cold, dense envelope and the 
desorption of molecular gas in the warmer outflow regions 
\citep{wiseman.et.al01}.

To investigate infall, the presence of other effects should be accounted for. 
This is not possible with one-dimensional radiative transfer codes because only
radial motions can be modelled. Two-dimensional codes are useful because many
systems will have cylindrical symmetry around the outflow axis. In fact the line profile observed from a molecular outflow changes dramatically as the angle varies between the outflow axis and the observers 
line of sight \citep{ward-thompson&buckley01}. To allow for arbitrary
viewing angle relative to the symmetry axis it is therefore necessary to
couple the 2D codes to 3D ray-tracing algorithms.
Specific sources may exhibit significant deviations from axisymmetry. These
could arise, for example, if the outflows are non-axisymmetric and/or there 
is significant non-symmetric structure in the envelope or circumstellar disk.
In these cases a fully three-dimensional approach is required. The drawback of
3D codes is that they have large memory requirements for fine grid resolution.

In this paper, we continue our coupled observational and modelling 
investigation of early stage collapse objects that are clearly undergoing 
dynamical activity and harbour a central accreting source. We analyse the 
emissions from a sample of three low mass
protostellar cores which possess evidence of both infall and outflows. 
The line profiles of several molecular line transitions are observed which 
effectively probe the conditions in the dense core, diffuse envelope and 
molecular outflow.
These are used to characterise the physical parameters (such as the 
temperatures, densities and chemical abundances) of each source in the context 
of a single unified dynamical model that includes infall and a bipolar 
outflow.
To do this we have utilized a 3D radiative transfer code, {\sc mollie} \cite[MOecular LIne Explorer,][] {keto.et.al04} that is specifically employed to
study how the line profiles are constrained by the orientation of the outflow
axis with respect to the observer.

We aim to show that a single unified model, comprising a spherically symmetric 
envelope/inflow and a bipolar outflow with a narrow interface region, can be 
used to describe the emission from each source.
We further wish to test the hypothesis that the significant source-to-source
variations that are observed are primarily the results of differences in the
source orientation/viewing angle.

Different isotopologues and transitions of the same molecular species preferentially trace somewhat physical different conditions within a dynamically active core. For example, an outflow may be best observed in a low excitation transition of an abundant transition such as $^{12}$CO or $^{13}$CO line whereas the envelope turbulent velocity may be better constrained by the hyperfine structure in, for example, C$^{17}$O lines. The source models were assembled by
firstly modelling the different dynamical components, using the transition in which they are most strongly apparent. Then, using this information, a self-consistent model for the entire source was constrained using all species and transitions.   

Section~\ref{sect:sources} summarises the properties of the sources
and Section~\ref{sect:obs} describes our continuum/molecular 
observations used to constrain the modelling.
Section~\ref{sect:modelling} introduces the {\sc mollie} code and our method, analysis, and the results obtained are given
in Section~\ref{sect:results}. Section~\ref{sect:discussion}
discusses these results and our concluding remarks are given in
Section~\ref{sect:conclusions}.

\section{Sources}
\label{sect:sources}

Our study concentrates on three protostellar cores; B335, I04166 and 
L1527, which are each believed to be in the Class 0 phase of evolution 
for low mass protostellar objects.
They have similar spectral energy distributions (SEDs) and, being of a 
similar evolutionary status, should be described by a single physical
model. 

Extensive observational data (molecular and continuum) and modelling 
efforts already exist for these sources and, significantly, 
spectral signatures of both infall and outflow are seen to be present.
We therefore hope to be able to identify the similarities and the 
source-to-source variations that exist within objects at a similar 
evolutionary stage.

\begin{figure*}
\centering
\includegraphics[width=5cm,height=15cm,angle=-90]{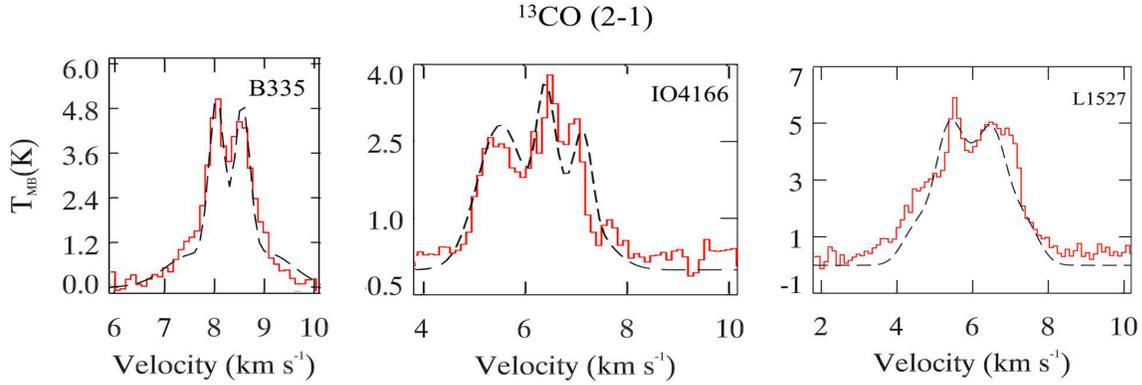}
\caption{$^{13}$CO ($J=2~$--$~1$) line profiles at the (0,0) offset 
position for B335, IO4166 \& L1527.}
\label{13co_compare}
\end{figure*}

As an example of the types of variations that are observed, 
Fig~\ref{13co_compare} presents the $^{13}$CO J=2-1 line profiles at the 
zero offset (0,0) position for each of the cores. Significant variations 
in profile shape and strength are clearly evident.

We briefly summarise here the key points from some previous studies, 
relating to the morphology and dynamical structure of each source.

\subsection{B335}
B335 is a very well studied core
\citep{hodapp98,wilner.et.al00,evans.et.al05,stutz.et.al08,yen.et.al10,yen.et.al11} located 250 pc away
\citep{tomita.et.al79} and contains the IRAS source $19347+0727$
\citep{huard.et.al99}. 

It is, to a large extent, the prototypical Class 0 infall source in 
that it was the first such object in which the classic asymmetric 
double-peaked infall line profiles were identified and analysed \citep{zhou.et.al93}. 
This analysis was helped by the facts that it is isolated, singular and 
approximately spherical. In addition, the outflow lobes are approximately 
in the plane of the sky; minimising the dynamical confusion between infall 
and outflows.

The infall component was detected through the observation of several 
transitions in both single dish and interferometric modes
\citep{chandler&sargent-93,zhou.et.al93,velusamy.et.al95,nisini.et.al99}.
The clear presence of infall, coupled with the existence of a massive 
envelope gas indicates significant accretion and is consistent with a 
Class 0 protostellar object \citep{andre93}.

Monte Carlo radiative transfer modelling by \cite{choi.et.al95}
was successful in describing the observed line emission towards the centre
of the molecular cloud in the context of the well-known spherically 
symmetric `inside out' collapse model of \cite{shu77}. 
An infall (`collapse expansion wave') radius of 0.035 pc 
was deduced \citep{zhou.et.al93}. 
The expected density profile within the free-fall zone ($\propto r^{-1.5}$)
is consistent with near-infrared extinction maps by
\cite{harvey.et.al01}. Outside the infall radius the density profile is 
consistent with that expected for a static singular isothermal sphere
\citep{saito.et.al99,shirley.et.al00}.

Emission from a bipolar molecular outflow was observed in
molecular line spectra by \cite{moriarty-schieven&snell89}. The
outflow is very close to the plane of the sky ($\sim$10$^{\circ}$)
and is elongated in the east-west direction
\citep{cabrit88,hirano88,saito.et.al99}. 

\subsection{I04166}
I04166 is a protostellar object located in the Taurus molecular
cloud at a distance of 140 pc \citep{elias78} and contains the IRAS source
$04166+2706$ \citep{kenyon-et-al90}. The radius of the molecular
gas cloud has been estimated from the extent of molecular line emission
from the high density tracers NH$_{3}$ (1,1) \& (2,2) and from 1.2
mm continuum emission. Both are coincident out to a distance of 0.03
pc \citep{tafalla-et-al04}.
\cite{motte-andre01} constructed a SED which is consistent with a Class 0 
source and observations of N$_{2}$H$^{+}$ ($J=1~$--$~0$) by \cite{chen-et-al07}
inferred a mean molecular hydrogen number density, 
$n_{\rm H_{2}} = 4 \times10^{6}$ cm$^{-3}$. 

I04166 contains a highly collimated molecular outflow. The axis
position angle (PA) is 30$^{\circ}$ (east of north) and extends
0.25 pc either side of the IRAS source \citep{bontemps-et-al96}.
\cite{tafalla-et-al04} observed the outflow in $^{12}$CO
($J=2~$--$~1$) emission and found there are two distinct velocity
components, one is at 50 km s$^{-1}$ and a slower component with a
velocity $<$ 10 km s$^{-1}$. A faster gas component is usually
associated with the youngest of protostellar objects
\citep{bachiller96}. The slow gas is believed to be created by 
a shear flow of accelerated ambient gas moving along the wall of
an evacuated cavity.

\subsection{L1527}
Like I04166, L1527 is a protostellar core located in the Taurus molecular
cloud at a distance of 140 pc and it contains the IRAS source 04368+2557
\citep{elias78}. Double peaked asymmetric line spectra consistent
with gravitationally infalling gas was observed by
\cite{zhou.et.al94} and \cite{myers.et.al95}. Observations by
\cite{zhou-et-al96} and radiative transfer modelling by
\cite{myers.et.al95} determined an infall radius of 0.03 pc.
Continuum observations from $100 \rightarrow 800$ $\mu$m shows the
gas density profile varies as $r^{-1.5}$ within a radius of
0.03 pc. 
An upper limit to the age of
this source was calculated by \cite{ohashi97} to be 10$^{5}$ yrs on the 
assumption of a constant accretion rate. However, \cite{kenyon-et-al93} 
found that a fit to the SED gives a lower age of 4.6 $\times$ 10$^{3}$
years indicating this is a young Class 0 source \citep{andre93}.

A molecular outflow was observed in single dish and
interferometric observations by \cite{macleod-et-al94} and
\cite{tamura-et-al96}. The outflow emission has an hourglass
morphology \citep{zhou-et-al96,ohashi97} with the outflow axis at
$<$ 10$^{\circ}$ to the plane of the sky \citep{tamura-et-al96}.

More recent {\em Spitzer} observations of L1527 have clearly revealed 
the bipolar outflow structure but have also identifed the presence of a 
`dark lane' which has been ascribed to a modified inner envelope/outflow
cavity morphology \citep{tobin.et.al08}. This inner `nozzle' has a distinct
effect on the {\em Spitzer} and near-infrared images of the source, but the
scale of the structure ($\sim$100\,au) is such that it would not significantly
affect line profiles as observed at single-dish resolution.   

\section{Observations and Data}
\label{sect:obs}

Line observations of our sources were made with the James Clerk
Maxwell Telescope (JCMT) over a period of 18 months (mostly in service mode)
using heterodyne receivers. We obtained a variety of strip and five-point maps
for the various sources/transitions that we have considered, which have given
sufficient data coverage of our sources so as to allow accurate modelling.
We have also used JCMT-SCUBA (the Submillimeter Common-User Bolometer Array)
archival data for 450/850$\mu$m continuum emission.

\begin{table*}
\centering \caption{The observed line emission tracers modelled in this paper.} 
\label{tab:obs}
\begin{tabular}{@{}lccccc@{}}
\hline\\[-1.0em] Source  & Molecule  & Transition & $n_{\rm crit}$ cm$^{-3}$ &
Receiver & Date Observed \\
\hline\\[-1.0em]
B335 & H$^{13}$CO$^{+}$& ($J=3~$--$~2$) & $3.4 \times 10^{6}$ & A3 & 20 - August - 2001\\
     & HCO$^{+}$& ($J=3~$--$~2$) & $1.8 \times 10^{6}$ & A2 & 18 -February - 1997\\
     & $^{13}$CO& ($J=2~$--$~1$) & $9.6 \times 10^{3}$ & A3 & 15 - May - 2003 \\
     & C$^{18}$O& ($J=2~$--$~1$) & $9.5 \times 10^{3}$ & A3 & 3 - May - 2003 \\
     & C$^{17}$O& ($J=2~$--$~1$) & $1.0 \times 10^{4}$ & A3 & 3 - May - 2003 \\
\hline\\[-1.0em]
I04166 & $^{13}$CO& ($J=2~$--$~1$) & $9.6 \times 10^{3}$ & A3 & 1 - April - 2003 \\
       & C$^{18}$O& ($J=2~$--$~1$) & $9.5 \times 10^{3}$ & A3 & 18 - March - 2003 \\
       & C$^{17}$O& ($J=2~$--$~1$) & $1.0 \times 10^{4}$ & A3 & 3 - May - 2003 \\
\hline\\[-1.0em]
L1527 & H$^{13}$CO$^{+}$& ($J=3~$--$~2$) & $3.4 \times 10^{6}$ & A3 & 3 - May - 2003 \\
      & $^{12}$CO& ($J=2~$--$~1$) & $1.1 \times 10^{4}$ & A3 & 3 - May - 2003 \\
      & $^{13}$CO& ($J=2~$--$~1$) & $9.6 \times 10^{3}$ & A3 & 3 - May - 2003 \\
      & C$^{18}$O& ($J=2~$--$~1$) & $9.5 \times 10^{3}$ & A3 & 3 - May - 2003 \\
      & C$^{17}$O& ($J=2~$--$~1$) & $1.0 \times 10^{4}$ & A3 & 3 - May - 2003 \\
\hline\\[-1.0em]

\end{tabular}
\end{table*}

\begin{table*}
\caption{Flux Calibration Factors from photometry observations of
Uranus. The calibration factors listed  are determined for 
40$^{\prime\prime}$ and 120$^{\prime\prime}$ beams. Emission maps
at $450~\mu{\rm m}$ and $850~\mu{\rm m}$ of B335, L1527 and 04166
were originally published in \citet{shirley.et.al00}.}
\label{tab:dust-calibrations}
\begin{center}
\begin{tabular}{@{}cccccccccc@{}}
\hline\\[-1.0em]
Source & $\alpha$ (J2000)& $\delta$ (J2000) & $\tau_{850}$ & $\tau_{450}$ & Date & 
FCF$^{850}_{40}$ & FCF$^{850}_{120}$ & FCF$^{450}_{40}$ & FCF$^{450}_{120}$\\
 & & & (Jy V$^{-1}$) & (Jy V$^{-1}$) & (Jy V$^{-1}$) & (Jy V$^{-1}$)\\
\hline\\[-1.0em]
B335 & 19 37 01.13 & 07 34 10.9 & 0.14 & 0.69 & 17-April-1998 & 1.01 & 0.86 & 5.84 & 3.98\\
I04166 & 04 19 42.5 & 27 13 36 & 0.17 & 0.77 & 30-August-1998 & 0.98 & 0.77 & 5.32 & 3.74\\
L1527 & 04 39 53.89 & 26 03 10.5 &0.15 & 0.76 & 24-January-1998 & 1.02 & 0.92 & 5.76 & 5.34\\
\hline
\end{tabular}
\end{center}
\end{table*}

\subsection{Molecular Line Observations}
The cores were observed with RxA3 (and RxA2) in August 2001 and March-May
2003. 
All observations were made in frequency switching mode and the details of
the transitions are given in Table~\ref{tab:obs}. The data were reduced 
using the {\sc specx} software package and further
analysis was undertaken with the {\sc class} package of IRAM. For
the majority of the observations, the system temperature was
between 300 and 450 K. A main beam correction factor of 0.69 was
used to convert the antenna temperatures into T$_{\rm{MB}}$.
The critical density for thermalisation ($n_{\rm crit}$) for each transition is
also given in Table~\ref{tab:obs}.
These show the wide range of densities ($9\times 10^3-3\times 10^6$cm$^{-3}$)
that are probed by the chosen molecular tracers/transitions.

\subsection{Dust Continuum Observations}
Archival SCUBA data of our three sources was retrieved and reduced
using the {\sc surf} package. The data consists of dust continuum
emission observed at 850$\mu$m and 450$\mu$m. All our science
observations are 64-point jiggle maps with a 120$^{\prime\prime}$
chop throw. Wavelength dependent submillimeter extinction,
$\tau_{850}$ and $\tau_{450}$ were calculated from skydip
observations. Bolometers at the edge of the array and those with
excessive noise ($V_{\rm rms}~\geq~60~\rm{nV}$) were removed
before the data was rebinned to 0.5~$\theta_{\rm{MB}}$ per pixel
on a J2000 coordinate system. The SCUBA emission maps were
converted to Jansky units using FCFs calculated from photometry
observations of Uranus. The FCFs for each source are listed in
Table~\ref{tab:dust-calibrations} for each source.

\section{Modelling}
\label{sect:modelling}

\subsection{The Physical Model}

For the purpose of the radiative transfer modelling we have
used a simple dynamical model that is similar to that described in
\cite{rawlings.et.al04}. This is a multi-component physical 
construct that includes a static envelope of molecular gas whose 
density varies with radius from the centre of the gas cloud and a 
bipolar outflow. The temperatures in the envelope and the outflow interface
components are taken to be spatially invariant. This is obviously a significant
simplification but, as shown in \citet{tsamis.et.al08}, in the case of the low-J lines observed at low spatial resolution 
that we are modelling,
the line profiles are generally more sensitive to abundance variations and 
the gas dynamics than they are to the temperature profile. Obviously, this assumption breaks down in the case of smaller beams 
and/or higher levels of excitation.
It is assumed that the envelope and the outflow are dynamically decoupled from eachother as the interactions between the two are most likely limited to narrow
interface regions.

The envelope comprises the quasi-static
gas reservoir that surrounds a protostellar core and, initially, 
has a density structure that is described by the Lame - Emden
equation \citep{chandrasekhar67}. Bonner - Ebert (BE) spheres
provide a set of solutions for isolated, pressure-bound, spherically 
symmetric isothermal clouds in hydrostatic equilibrium and of varying 
density contrast between the centre and the edge of the cloud
\citep{bonner56,ebert55}.
The sources that we are studying have evolved beyond the initial 
stage of hydrostatic equilibrium and early collapse and we 
approximate the gas distribution with the 
simple Plummer density profile \citep{arreaga-garcia.et.al10}
\begin{equation}
\rho(r) = \frac{\rho_{0} R_{0}^{2}}{(R_{0}^{2} + r^{2})}
\label{eqn:plummer}
\end{equation}
where $\rho_{0}$ is the peak gas density and $R_{0}$ is the radius
within which the density is $\rho_{0}$. 
This profile is similar to the Bonnor - Ebert solution, retaining the 
$1/r^2$ shape at large radii, but has a somewhat flatter density 
distribution in the more central regions, as applicable to more evolved 
objects. The adoption of other representations of the density structure, such 
as simple power laws, or the Bonnor - Ebert solution, actually has very
little effect on the results.
This is because the differences between the profiles are greatest in the 
central regions which are poorly resolved and beam-diluted.

The outflow is driven by a high speed, low density hot jet which, through turbulent shear, accelerates the molecular gas. Within the outflow the temperature, velocity and density will vary with distance from the jet axis. Observations \citep[e.g.][]{gueth&guilloteau99} indicate that the highest velocity and most excited molecular gas is typically located in a narrow region close to the underlying jet. For convenience the molecular outflow is divided
into two regions, a hot, low density inner region which is close to the
protostellar jet (hereafter the `inner boundary layer') and a denser, cooler region
at the interface between the outflow envelope (hereafter the `outer boundary layer'). A better approximation would be a linear variation in the outflow parameters between the interface with the jet and with the envelope; however, due to the resolution of the single dish observations being modelled, this 
refinement would result in little change to the line profiles. So, for
simplicity, we emply a two-component molecular outflow model. 

\begin{figure}
\label{model}
\centering
\includegraphics[width = 8cm]{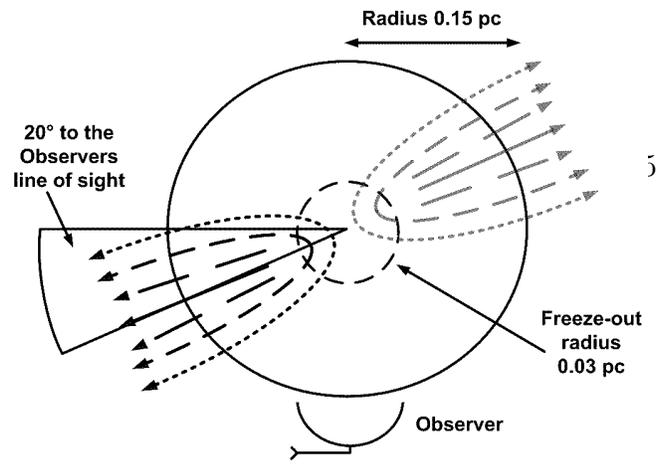}
\caption{Schematic diagram of the dynamical model. The dashed line is the outer boundary layer, the broken lines are the inner boundary layer and the solid line represents the unobserved underlying jet. The viewing angle (the tilt 
between the axis of the outflow lobes and the plane of the sky) and freeze-out
radius are for I04166. The dynamical models for L1527 and B335 have
different viewing angles and radii, recorded in 
Table~\ref{tab:best-fit-params-env}.} 
\label{schematic}
\end{figure}

Following \citet{rawlings.et.al04} and \citet{carolan.et.al08,carolan.et.al09} the shape of the molecular outflow is constructed using a {\em tanh} 
function to approximate to the observed
morphology of the base of the outflowing molecular gas. The
equation defining the edge of the outer boundary layer is,
\begin{equation}
z=\tanh(\lambda r),
\end{equation}
where $z$ and $r$ are the ordinates along and perpendicular to the outflow 
axis respectively.
Thus, $\lambda$ defines the shape of the flow. A value of $\lambda=2$ gives a morphology consistent with typical interferometric observations of low velocity outflows \citep[e.g.][]{lee.et.al02,arce&goodman02,jorgensen.et.al07}. 
The edge between the inner and outer boundary layers is defined by a value of
$\lambda=2.2$ \citep[as in][]{rawlings.et.al04,carolan.et.al08} so that
$\sim 10-20$ percent of the outflow is contained within the outer boundary 
layer. Within this layer, the velocity of the gas was
maintained at a constant speed with a direction tangential
to the inside edge of the boundary layer. Within the inner boundary layer, the velocity of the gas was maintained at a constant radial velocity.
For the sake of simplicity, in this model the density is taken to be constant 
within each of the boundary layers. More complex geometries (including density
and velocity gradients) were found not to have significant impacts on the 
line profiles.  Figure~\ref{schematic} is an example schematic diagram of the
model used for I04166.
Whilst this model is simplistic, it is consistent with the observational effects
that can be detected at single-dish resolution. Real outflow sources will have
more complex (sub-)structure 
\citep[e.g. as seen in L1527 -][]{tobin.et.al08} but on scales that will not
significantly affect the line profiles as observed at 
$\sim 10-20^{\prime\prime}$ resolution.

\subsection{Radiative Transfer Modelling}

In order to interpret the observed line emission profiles, a thorough 
analysis of the spectra must account for the simultaneous 
processes of infall, outflow, gas freeze-out and desorption including
the contributions to the profile formation at all points along the 
line of sight.

We have utilised a 3D radiative transfer code, {\sc mollie} (MOLecular LIne
Explorer)\footnote{https://www.cfa.harvard.edu/~eketo/mollie/documentation},
written and developed by Keto and collaborators
\citep[see, e.g.][for examples of its use]
{keto90,keto.et.al04,redman.et.al04a,redman.et.al06,
carolan.et.al08,carolan.et.al09,longmore.et.al11}. {\sc mollie} is used to generate synthetic line profiles to compare with observed rotational transition lines. In order to calculate the level populations the statistical equilibrium equations are solved using an Accelerated Lambda Iteration (ALI) algorithm \citep{rybicki&hummer91} that reduces the radiative transfer equations to a series of linear problems that are solved quickly even in optically thick conditions. {\sc mollie} splits the overall structure of a cloud into a 3D grid of distinct cells.

The input to {\sc mollie} is divided into voxels (3D pixels) and 
there are five input parameters which need to be uniquely defined 
for each voxel: the number density of H$_{2}$; the gas temperature; the gas bulk velocity; the microturbulent velocity of the gas; and the chemical abundance. 
As described in the next section, the parameters other than the abundance can each be constrained from the continuum and molecular observations, coupled with
radiative transfer calculations. 

\citet{carolan09} calculated the chemical abundances in a detailed chemical
model for simple spherically symmetric clouds and investigated the sensitivity
of the resulting molecular line profiles to the chemical variations. He found
that, because the largest chemical changes tend to take place in the centre of
the cloud (as a result of molecular freeze-out), the CO abundance in the 
bulk of the volume of the cloud can be taken to be approximately constant. 
Again, we note that the line profiles convolved to the resolution of single
dish observations do not significantly resolve the chemical variations. Therefore, we have not performed any chemical modelling in this study, rather 
we use the observations to constrain the molecular abundance for each of the
envelope, inner boundary layer and outer boundary layer in the sources.

Subsequently we perform a chi-squared analysis
to find the best fit parameters. This procedure works by running a
series of simulations with the abundance kept constant whilst the other four
parameters are varied. This is repeated for different abundances
until a best fit is found \citep{carolan09}.

\section{Analysis and Results}
\label{sect:results}

The chi-squared analysis described above can be used to obtain best fit
parameters from an arbitrary first approximation but, wherever feasible, we 
try to simplify the procedure by using simple analyses of observational data 
to constrain as many of the free parameters as possible. These values are then
refined by the modelling/chi-squared fitting analysis.

In practice, this is not possible for the outflow and interface components.
However, in the case of the envelope, if we make the assumptions described 
above (spherical symmetry, density structure described by a Plummer law, and
isothermality) we can make some rough first estimates of the key physical 
parameters, using optically thin C$^{18}$O $(J = 2~$--$~1)$ line and/or
dust continuum emission data.

Using this approach we obtain constraints on the following quantities: 
(a) the characteristic radius within which the CO abundance is significantly 
reduced as a result of freeze-out onto dust grains, by (b) a representative
depletion factor, (c) the normalisation ($\rho_0$,$R_0$) for the (Plummer)
density profile, (d) the gas temperature, and (e) the turbulent velocity.
The (highly simplified) procedure that we adopt is as follows:

CO freezes out onto the surface of dust grains at low temperatures 
($T \la 20$K) and high densities ($n_{\rm{H}_{2}} \ga 10^{4}$ cm$^{-3}$) 
\citep{sandford}. To determine the degree of freeze-out we compare spatial 
distributions of the H$_2$ column density as determined from the dust 
continuum observations to those determined from molecular (CO) observations. 
The offset between the two gives a measure of the amount of molecular 
depletion on the assumption that freeze-out is the the only cause of the CO
abundance reduction. 
The absolute normalisation will depend on such factors as the dust opacity, 
the dust-to-gas ratio and the (assumed constant) CO$:$H$_2$ ratio, and so is
highly uncertain, but the relative variations across the core will be
reasonably well-defined.

The column density of H$_{2}$ can be derived from the dust continuum fluxes
using the equation
\begin{equation}
N_{\textrm{H}_{2}} = \frac{S_{850}}{\Omega \kappa_{850} \mu
m_{\rm{H}} B(\rm{T_{dust}})},
\label{column-den-from-dust}
\end{equation}
where $m_{\rm{H}}$ is the mass of a hydrogen atom and $\mu$ is the 
mean molecular weight.
$S_{\rm{850}}$ is the flux in Jy at 850$\mu$m, 
$\Omega$ is the aperture solid angle in steradians;
$\Omega = (\pi \Theta^{2})/(4 \ln 2)$, where $\Theta$ is the beam
size in radians. We have measured the flux in a 21$^{\prime\prime}$ beam
so that a suitable comparison to the C$^{18}$O $J=2-1$ observations 
(made with a HPBW of 21$^{\prime\prime}$) can be made.  The Planck function $B(\rm {T_{dust}})$ is calculated assuming a 
uniform dust temperature of 10K. This is a simplification, but
a number of other studies \citep[e.g.][]{evans.et.al05}
have shown a beam-averaged temperature of $\sim$10K is
appropriate for Class 0 sources. There will obviously be some 
deviations from this value near the outer edge of the envelope, and close to
the protostar but, at single dish resolution we find that the effects are
marginal.
A dust mass opacity of $\kappa_{850}$ = 0.02 cm$^{2}$~g$^{-1}$ is 
assumed which is consistent with \cite{ossenkopf&henning94} - being
representative of dust grains with thin ice mantles. The dust to gas
ratio is taken to be 100.

\begin{figure*}
\centering
\includegraphics[width = 6cm, height=5cm]{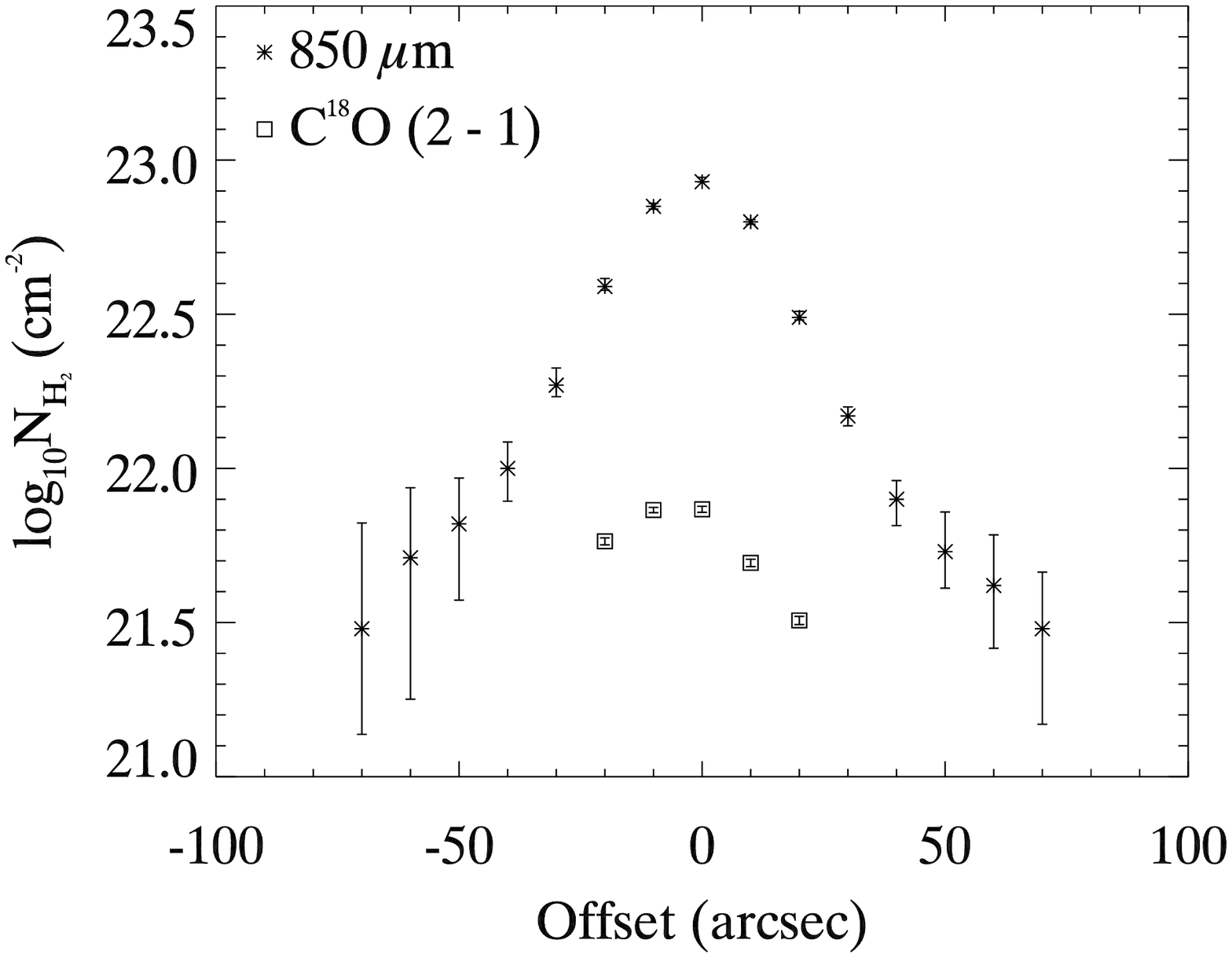}
\includegraphics[width=6cm]{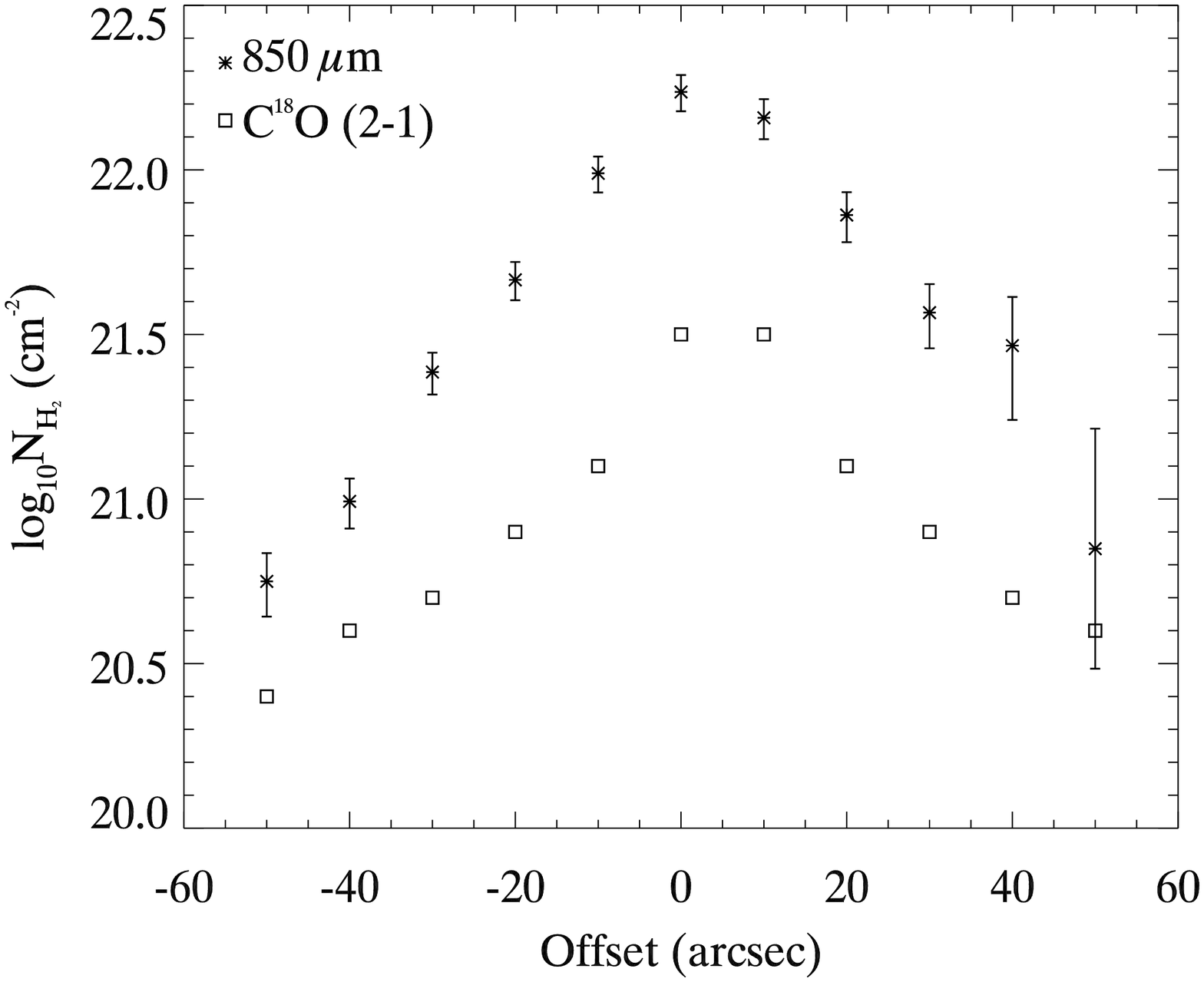}
\includegraphics[width=6cm]{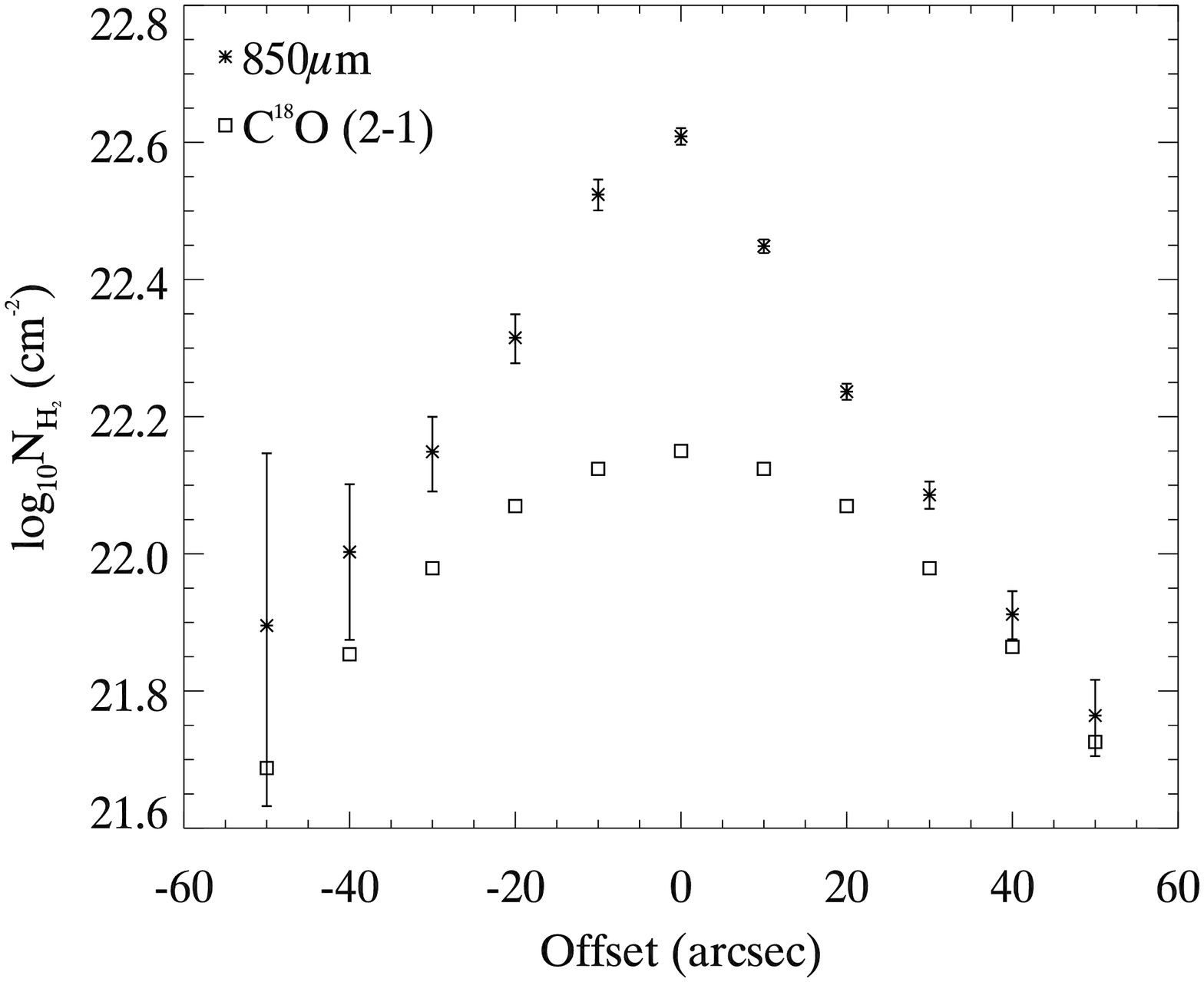}
\caption{A comparison of the column density of H$_{2}$ derived
from SCUBA dust continuum and integrated C$^{18}$O molecular line
emission for B335, I04166 and L1527 respectively.}
\label{fig:freezeout}
\end{figure*}

The column density of H$_{2}$ has also been determined from molecular gas
observations using 
the optically thin rotational transition C$^{18}$O $(J = 2~$--$~1)$, with 
the simplifying assumptions of isothermality and a standard CO$:$H$_2$ ratio.
This analysis is fully described in Appendix A. 

Figure~\ref{fig:freezeout} shows comparisons between the two as observed in 
each of the three sources. The dust emission profiles closely match the 
results from previous studies \citep[e.g.][]{shirley.et.al02}.
Each plot in Figure~\ref{fig:freezeout} clearly shows the presence of
depletion, which becomes more significant as one approaches the centre of 
each core.
We simplify the complexity of this structure by a step function, so that a 
nominal depletion radius is defined for each source, within which the gas-phase
CO abundance is reduced by a factor of between 3 and 10. These parameters are 
initially estimated from Figure~\ref{fig:freezeout} and then futher refined by
the modelling/fitting technique.

Using a Plummer law for the density profile and assuming spherical symmetry
in the envelope we then use the column densities of H$_{2}$ derived from the (undepleted) dust continuum emission to determine the normalisation factors 
($R_0$, $\rho_0$) in equation~\ref{eqn:plummer} and hence the number density 
of H$_{2}$: $n_{H_{2}}$. 

The dust temperature was found by first re-binning the 
450$\mu$m emission to 14$^{\prime\prime}$ which is the size of the beam at
850$\mu$m. A flux density is then extracted which is the azimuthal
average of the flux density in a 14$^{\prime\prime}$ aperture. The
flux density was measured at 7$^{\prime\prime}$ spacing to ensure that
the emission is Nyquist sampled. Note that this assumes a perfect (gaussian) beam. In practice, we find 
that the inclusion of the error beams at 450 and 850$\mu$m makes very 
little difference to this analysis.
The temperature can then be estimated from the ratio of the fluxes at
450 and 850$\mu$m. At these wavelengths there are significant departures from
the Rayleigh-Jeans approximation and the flux ratio is given by;
\begin{equation}
\frac{S_{450}}{S_{850}} = \left(\frac{850}{450}\right)^{3+\beta}
\frac{\exp(17K/T_{\rm dust})-1}{\exp(32K/T_{\rm dust})-1}
\label{eqn:color-diff}
\end{equation}
\citep{kramer.et.al03a}, where $S$ is the flux in a 14$^{\prime\prime}$ beam, 
$T_{\rm dust}$ is the dust temperature and $\beta$ is the dust emissivity
index. For the purposes of our analysis we use $\beta$ = 1.5 which is
characteristic of dust grains in cold molecular clouds, is reasonably 
consistent with \cite{ossenkopf&henning94} and is within the limits of 
variation ($1\leq\beta\leq2$) expected for molecular clouds 
\citep{schnee&goodman05}.

There are a couple of major assumptions with this method; (i) We assume that 
the dust and gas are well-coupled, in which case a single value can be used 
to describe both the gas and the dust temperatures. For the dense cores that 
we are investigating this us a fair approximation. (ii) We do not allow for any 
spatial variation in temperature across the envelope. This is known not to 
the case \citep[e.g.][]{shirley.et.al02} but the magnitude of those variations
are not sufficient to warrant a more detailed approach in a simple model that
is tailored to what can be detected at single-dish resolution.
 
Finally, the turbulent velocity, $\sigma_{\rm NT}$, is estimated on the
assumption that turbulence is the sole non-thermal component of the line
broadening for the (optically thin) single peaked C$^{18}$O $(J = 2~$--$~1)$
lines. In which case;
\begin{equation}
\sigma_{\rm tot}^2 = \sigma_{\rm NT}^2 + \sigma_{\rm T}^2,
\label{eqn:turb-linewidth}
\end{equation}
where $\sigma_{\rm tot}$ is the total velocity dispersion and is related to
the observed FWHM of a representative molecular line profile ($\Delta\upsilon$)
by $\sigma_{\rm tot}^2 = \Delta\upsilon/\sqrt{8 \ln 2}$.
$\sigma_{\rm T}$ is the thermal velocity dispersion given by 
$\sigma_{\rm T}^2 = kT/\mu$, where $\mu$ is the molecular mass, $T$ is the gas
temperature and $k$ is Boltzmann's constant. 

\subsection{Radiative Transfer Modelling of the Observed Spectral
Line Emission} 

Our results and best-fit parameters are described in this section 
for each of our three sources.
Table~\ref{tab:obs} lists the molecular transition observed for each source 
and the results are presented in figures~\ref{b335_c17o}-\ref{l1527_12co}
with the observed emission line profiles overlayed with the best fit modelled
profiles. The best fit model parameters are given in 
tables~\ref{tab:best-fit-params-env}-\ref{tab:best-fit-params-inner-bnd}.
The number of significant figures given in these tables indicates the level of
accuracy obtained by the chi-squared fitting technique. 

\subsubsection{B335}

\begin{figure*}
\centering
\includegraphics[width = 15cm]{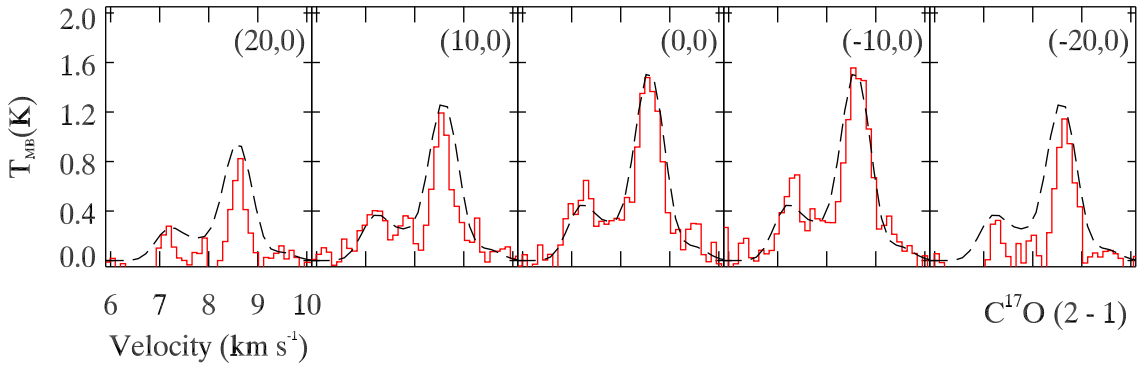}
\caption{B335: C$^{17}$O ($J=2~$--$~1$) line profiles:
observed (solid line) and modelled (dashed line).
The offset between cells is 10$^{\prime\prime}$.} 
\label{b335_c17o}
\end{figure*}

\begin{figure*}
\centering
\includegraphics[width = 15cm]{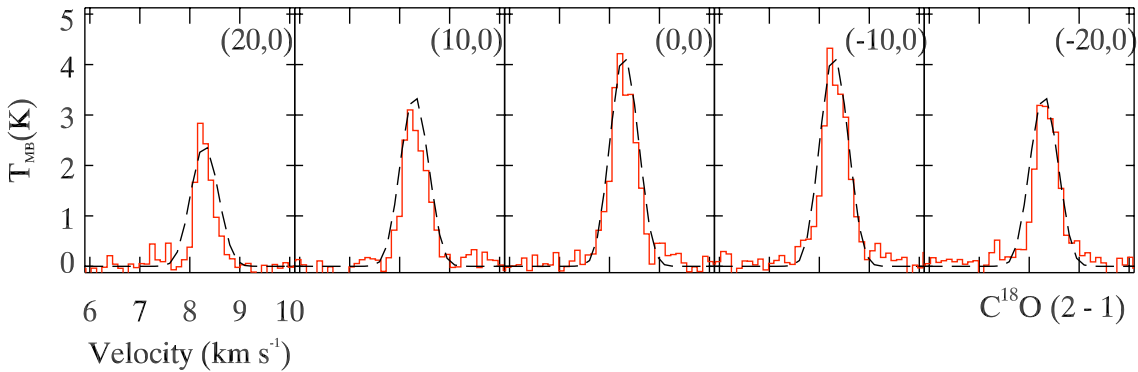}
\caption{B335: C$^{18}$O ($J=2~$--$~1$) line profiles:
observed (solid line) and modelled (dashed line).
The offset between cells is 10$^{\prime\prime}$.} 
\label{b335_c18o}
\end{figure*}

B335 was observed in several molecular species/transitions. Line
profiles of the optically thin ($J=2~$--$~1$) transition of
C$^{17}$O and C$^{18}$O are shown in Figures~\ref{b335_c17o}
\&~\ref{b335_c18o}. They are shown as a strip of spectral line
profiles that are centered at the peak of 850$\mu$m dust continuum
emission which corresponds to the (0,0) offset in both
Figures~\ref{b335_c17o} and \ref{b335_c18o}.

The observed line emission profile of C$^{17}$O ($J=2~$--$~1$)
arises from separate hyperfine line transitions whose peaks are
clearly distinguishable in Figure~\ref{b335_c17o}. In a highly
turbulent gas the components would be blended. The spectra of 
C$^{18}$O ($J=2~$--$~1$) are single peaked and have similar intrinsic line
widths as the C$^{17}$O ($J=2~$--$~1$) lines. It would therefore seem that
both transitions are preferentially probing cold, quiescent gas in the envelope.
The best fit parameters from the radiative transfer modelling
of the molecular gas envelope are listed in
Table~\ref{tab:best-fit-params-env}.
Blanks in the table imply that the line profiles are insensitive to the 
abundances which are therefore unconstrained by the model.

The turbulent velocity in the envelope gas was calculated to be 
$0.25$ km s$^{-1}$ as determined from the C$^{18}$O ($J=2~$--$~1$) line 
profiles. To account for the observed molecular depletion, as evident in 
Figure~\ref{fig:freezeout}, the C$^{18}$O ($J=2~$--$~1$) and 
C$^{17}$O ($J=2~$--$~1$) line profiles were modelled with an abundance that
is reduced by a factor of ten inside a freeze-out radius of $\sim 0.3\times$ 
the core radius.

The best fit abundances of C$^{18}$O obtained inside and outside the freeze-out
region are $3 \times 10^{-8}$ and $3 \times 10^{-7}$ respectively.
These are similar values to those obtained by \cite{evans.et.al05}:
$2.5 \times 10^{-8}$/$7.4 \times 10^{-8}$.
C$^{17}$O was modelled with the same depletion factor so that the best
fit abundances were $0.35 \times 10^{-8}$/$3.5 \times 10^{-8}$. 
This implies a value for the $^{18}$O$:^{17}$O ratio that is a factor of 
$\sim 2\times$ larger than interstellar values \citep{schoier.et.al02}.

\begin{figure}
\centering
\includegraphics[width = 4cm]{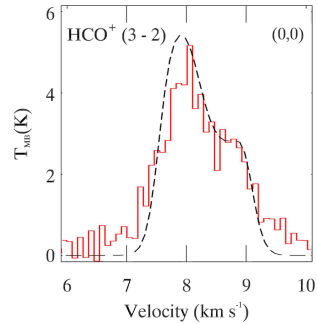}
\caption{B335: HCO$^{+}$ ($J=3~$--$~2$) line profiles:
observed (solid line) and modelled (dashed line).
The middle panel is at 0$^{\prime\prime}$ offset.} 
\label{b335_hcoplus}
\end{figure}

\begin{figure}
\centering
\includegraphics[width = 8cm,angle=90]{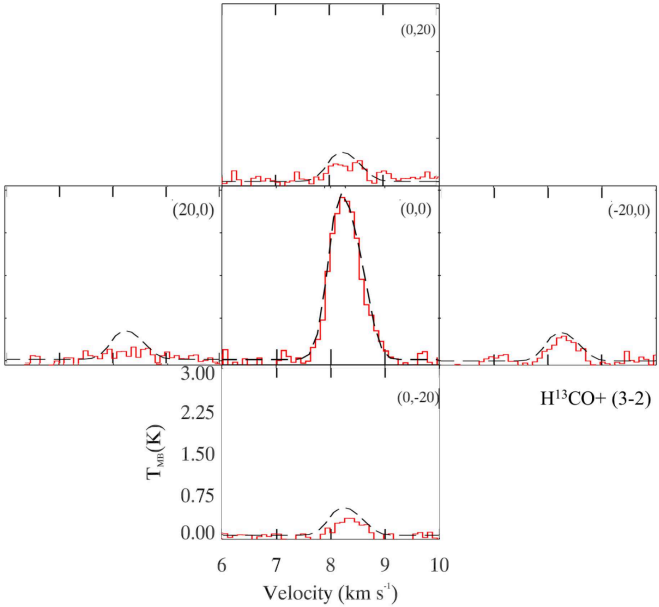}
\caption{B335: H$^{13}$CO$^{+}$ ($J=3~$--$~2$) line profiles:
observed (solid line) and modelled (dashed line).
The offset between cells is 20$^{\prime\prime}$ and
the middle panel is at 0$^{\prime\prime}$ offset.} 
\label{b335_h13coplus}
\end{figure}

Figure~\ref{b335_hcoplus} shows the observed line spectra and the
modelled profiles for HCO$^{+}$ ($J=3~$--$~2$). 
The observational data consists of a single line profile at zero offset.
H$^{13}$CO$^{+}$ ($J=3~$--$~2$) line profiles are shown in 
Figure~\ref{b335_h13coplus}. Although clearly present in the zero offset 
position, there is barely any emission detected in the (20$^{\prime\prime}$) 
off centre spectra. 
This is not surprising as the critical densities of the transitions are 
1.8 $\times$ 10$^{6}$ cm$^{-3}$ and 3.4 $\times$ 10$^{6}$ cm$^{-3}$ 
respectively; larger than the peak H$_{2}$ density in B335.

The values of the temperature, peak density, infall and turbulent velocities
were constrained from the C$^{18}$O ($J=2~$--$~1$) \& C$^{17}$O ($J=2~$--$~1$)
lines. The best fit HCO$^{+}$ abundance is $3 \times 10^{-8}$ which agrees 
well with the value ($3.5 \times 10^{-8}$) obtained by \cite{evans.et.al05}. 
The H$^{13}$CO$^{+}$ best fit abundance is $9 \times 10^{-10}$. The ratio
H$^{12}$CO$^{+}$/H$^{13}$CO$^{+}$ is $\approx$ 39. This is less
than the standard interstellar abundance ratio $^{12}$C/$^{13}$C
$\sim 70$ measured in atomic clouds \citep{wilson&rood94} and may be indicative of chemical fractionation effects as discussed below.

\begin{figure*}
\centering
\includegraphics[width = 15cm]{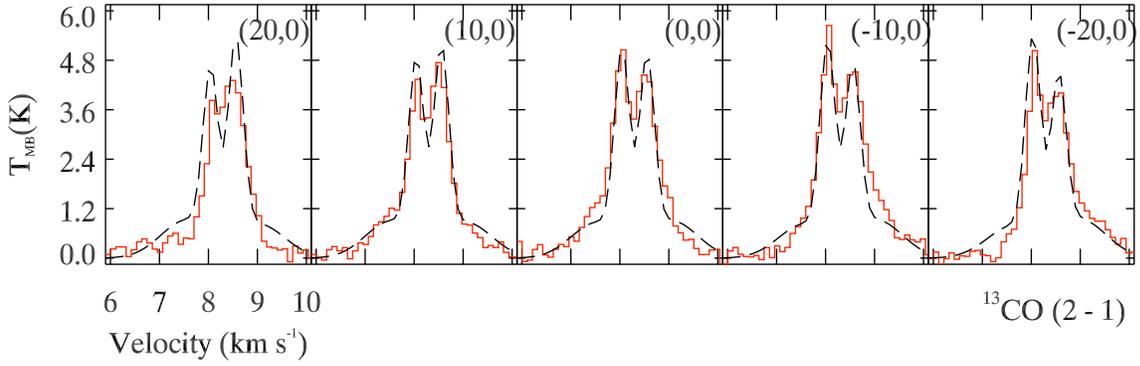}
\caption{B335: $^{13}$CO ($J=2~$--$~1$) line profiles:
observed (solid line) and modelled (dashed line).
The offset between cells is 10$^{\prime\prime}$.}
\label{b335_13co}
\end{figure*}

Figure~\ref{b335_13co} shows the observed and modelled line
emission spectrum for $^{13}$CO ($J=2~$--$~1$). 
The line profiles are double peaked and asymmetric. The critical density 
for this transition is three orders of magnitude less than the peak 
density in B335 so that it probes the optically thick gas far from the
centre of the cloud. The asymmetry changes from red-asymmetric
(i.e. red$>$blue) at (20,0) to blue-asymmetric at (-20,0).

Asymmetric, double peaked line profiles are often taken to be signatures
of the presence of infall. However, such an explanation fails to explain
the facts that (a) both red and blue-asymmetry are present, and (b) 
asymmetric emission is present at significantly greater offsets
($>10^{\prime\prime}$) than expected for an infall source. 
A bipolar outflow is therefore a much more likely cause and naturally 
gives rise to spatially distinct redshifted and blueshifted components.
Moreover, outflowing gas has been observed in $^{12}$CO ($J=2~$--$~1$) 
by \cite{moriarty-schieven&snell89} extending to 0.36 pc
(5$^{\prime}$) either side of IRAS 19347+0727. The emission is
observed to increase in intensity at the lobe edge.
The asymmetry in the line emission profile is essentially caused by
the axis of outflow axis being tilted to the line of sight. 
Tables~\ref{tab:best-fit-params-outer-bnd}
\&~\ref{tab:best-fit-params-inner-bnd} specify the the best fit parameters 
for the gas in the outer and inner boundary layer of the outflow.

\begin{table*}
\centering \caption{Best fit parameters for the gas in the
envelope of each source. The first eight parameters were roughly
constrained by observations and then fine tuned to improve the model
fits. The outflow tilt angle is relative to the plane of the sky and
the abundance is the column density ratio relative H$_{2}$.}
\label{tab:best-fit-params-env}
\begin{tabular}{@{}llccc@{}}
\hline\\[-1.0em]
Envelope Parameters & & B335 & I04166 & L1527 \\
\hline\\[-1.0em]
Outflow tilt angle (degrees) & & -10 & 20 & 10  \\
Source radius & (pc) & 0.15 pc & 0.15 pc & 0.08 pc  \\
Depletion radius & (pc) & 0.05 & 0.03 & 0.015 \\
CO depletion factor & & 10 & 3 &3 \\
Temperature& (K) & 20 & 20 & 20 \\
Peak Density & (cm$^{-3}$) & 10$^{6}$ & $4\times 10^6$ & 10$^{6}$ \\
Velocity & (km s$^{-1})$ & -0.3 & -0.2 & -0.3 \\
Turbulent width & (km s$^{-1})$ & 0.25 & 0.2 & 0.3 \\
Abundance C$^{17}$O &($\times 10^{-8}$) & 3.5 & 0.98& 0.075 \\
Abundance C$^{18}$O &($\times 10^{-8}$) & 30 & 2.6 & 0.30 \\
Abundance $^{13}$CO &($\times 10^{-8}$) & 180 & 6 & 5 \\
Abundance $^{12}$CO &($\times 10^{-8}$) & & & 400 \\
Abundance HCO$^{+}$ &($\times 10^{-8}$) & 3 & & \\
Abundance H$^{13}$CO$^{+}$ &($\times 10^{-8}$) & 0.09 & & 0.0022 \\
 \hline
\end{tabular}
\end{table*}

\begin{table*}
\centering \caption{Best fit parameters for the gas in the outer
boundary layer. This corresponds to the molecular gas at
the interface between the molecular outflow and the envelope. The
first four parameters were constrained by observations and then fine
tuned to improve the model fits. The abundances are relative to H$_2$.}
\label{tab:best-fit-params-outer-bnd}
\begin{tabular}{@{}llccc@{}}
\hline\\[-1.0em]
Parameters Outer & & B335 & I04166 & L1527 \\
Boundary Layer & & & & \\
\hline\\[-1.0em]
Temperature& (K) & 35 & 50 & 30 \\
Peak Density& (cm$^{-3}$) & $10^5$ & $10^5$ & $9 \times 10^4$ \\
Velocity  &(km s$^{-1})$ & 1.4 & 1.2 & 1.3 \\
Turbulent width & (km s$^{-1})$ & 0.3 & 0.4 & 0.4 \\
Abundance $^{13}$CO &($\times 10^{-8}$) & 300 & 10 & 60  \\
Abundance $^{12}$CO &($\times 10^{-8}$) & & & 450 \\
Abundance HCO$^{+}$ &($\times 10^{-8}$) & 9 & & \\
\hline
\end{tabular}
\end{table*}

\begin{table*}
\centering \caption{Best fit parameters for the gas in the inner
boundary layer of each source. The inner boundary layer is the
molecular gas in the outflow that is closest to the protostellar
jet. The first four parameters were constrained by observations and fine 
tuned to improve the model fits. The abundances are relative to H$_2$.}
\label{tab:best-fit-params-inner-bnd}
\begin{tabular}{@{}llccc@{}}
\hline\\[-1.0em]
Parameters Inner & & B335 & I04166 & L1527 \\
Boundary Layer & & & & \\
\hline\\[-1.0em]
Temperature& (K) & 55 & 80 & 60 \\
Peak Density& (cm$^{-3}$) & $10^4$ & $5\times 10^4$ & $10^4$ \\
Velocity  &(km s$^{-1})$ & 3.5 & 1.6 & 2.5 \\
Turbulent width &(km s$^{-1})$ & 0.5 & 0.6 & 0.6 \\
Abundance $^{13}$CO &($\times 10^{-8}$) & 600 & 12 & 400 \\
Abundance $^{12}$CO &($\times 10^{-8}$) & & & 600 \\
Abundance HCO$^{+}$ &($\times 10^{-8}$) & 12 & & \\
 \hline
\end{tabular}
\end{table*}

\subsubsection{I04166}

\begin{figure}
\centering
\includegraphics[width = 8cm,angle=90]{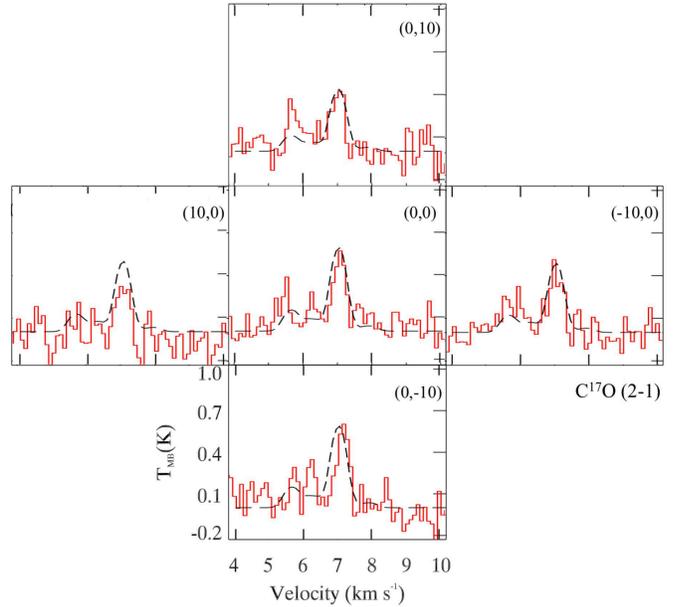}
\caption{I04166: C$^{17}$O line profiles:
observed (solid line) and modelled (dashed line).
The offset between cells is 10$^{\prime\prime}$.} 
\label{04166-c17o}
\end{figure}

\begin{figure}
\centering
\includegraphics[width = 8cm,angle=90]{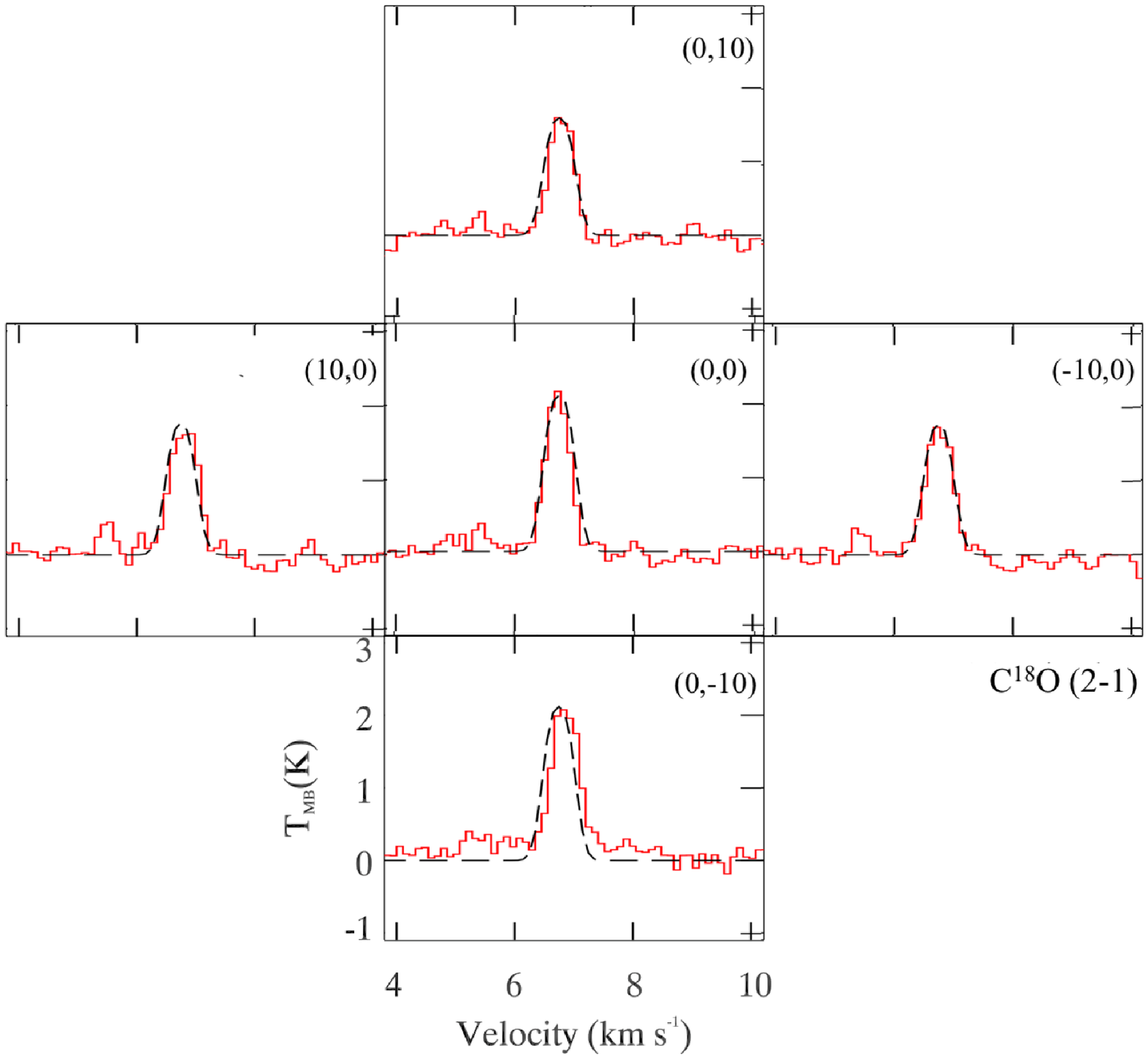}
\caption{I04166: C$^{18}$O line profiles:
observed (solid line) and modelled (dashed line).
The offset between cells is 10$^{\prime\prime}$.} 
\label{04166-c18o}
\end{figure}

Observations of C$^{17}$O and C$^{18}$O ($J=2~$--$~1$) towards I04166 
are shown in Figures~\ref{04166-c17o} and \ref{04166-c18o}. The line
profiles shown in Figure~\ref{04166-c17o} clearly show the
C$^{17}$O ($J=2~$--$~1$) hyperfine line transitions whilst the 
C$^{18}$O ($J=2~$--$~1$) lines shown in Figure~\ref{04166-c18o} are 
single-peaked and narrow. Both are characteristic of the quasi-static/slowly
infalling gas in the molecular gas envelope. 
Best fit parameters for these profiles (applying to the molecular 
envelope) are given in Table~\ref{tab:best-fit-params-env}. There is little
contribution to the modelled line profile from gas residing in the
molecular outflow. We also note that the peak density, as obtained
from the 850$\mu$m continuum observations is consistent with previous
estimates derived from N$_{2}$H$^{+}$ ($J=1~$--$~0$) observations
\citep{chen-et-al07}.

\begin{figure}
\centering
\includegraphics[width = 8cm,angle=90]{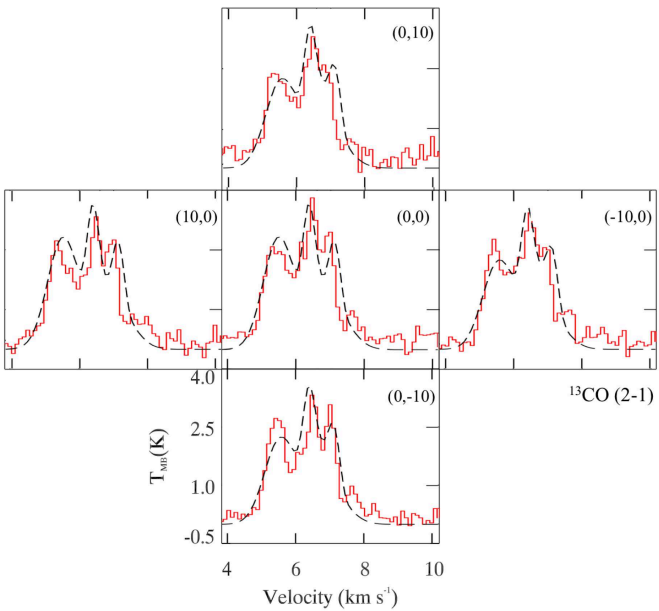}
\caption{I04166: $^{13}$CO ($J=2~$--$~1$) line profiles:
observed (solid line) and modelled (dashed line).
The offset between cells is 10$^{\prime\prime}$.}
\label{04166_13co}
\end{figure}

The line profiles for $^{13}$CO ($J=2~$--$~1$) are shown in 
Figure~\ref{04166_13co}. This shows the standard blue-asymmetric 
double-peaked line profile suggestive of infalling gas.
In addition there is a component at 5.5 km s$^{-1}$ which is also seen in
the optically thin C$^{18}$O ($J=2~$--$~1$) and, to a
lesser extent, C$^{17}$O ($J=2~$--$~1$) line profiles. It has also been
detected in the line profiles of CS ($J=2~$--$~1$) and H$_{2}$CO 
(2$_{12}$ - 1$_{11}$) \citep{mardones.et.al97}.

The best fit parameters for the infalling envelope component are given
in Table~\ref{tab:best-fit-params-env}. 
The high velocity gas component at 5.5 km s$^{-1}$ can be explained
by emission originating from the outer boundary layer with
the parameters given in Table~\ref{tab:best-fit-params-outer-bnd}. 
The wings of the line profile are explained by the high velocity warm 
gas in the inner boundary layer that is close to the protostellar jet.
The best fit parameters for this gas are given in
Table~\ref{tab:best-fit-params-inner-bnd}. Note that the blueshifted 
emission seen in Figure~\ref{04166_13co} is dominant due to the greater 
amount of absorbing gas on the redshifted side 
\citep{tafalla-et-al04,santiago-garcia-et-al09}.

\subsubsection{L1527}
\begin{figure}
\centering
\includegraphics[width = 8 cm]{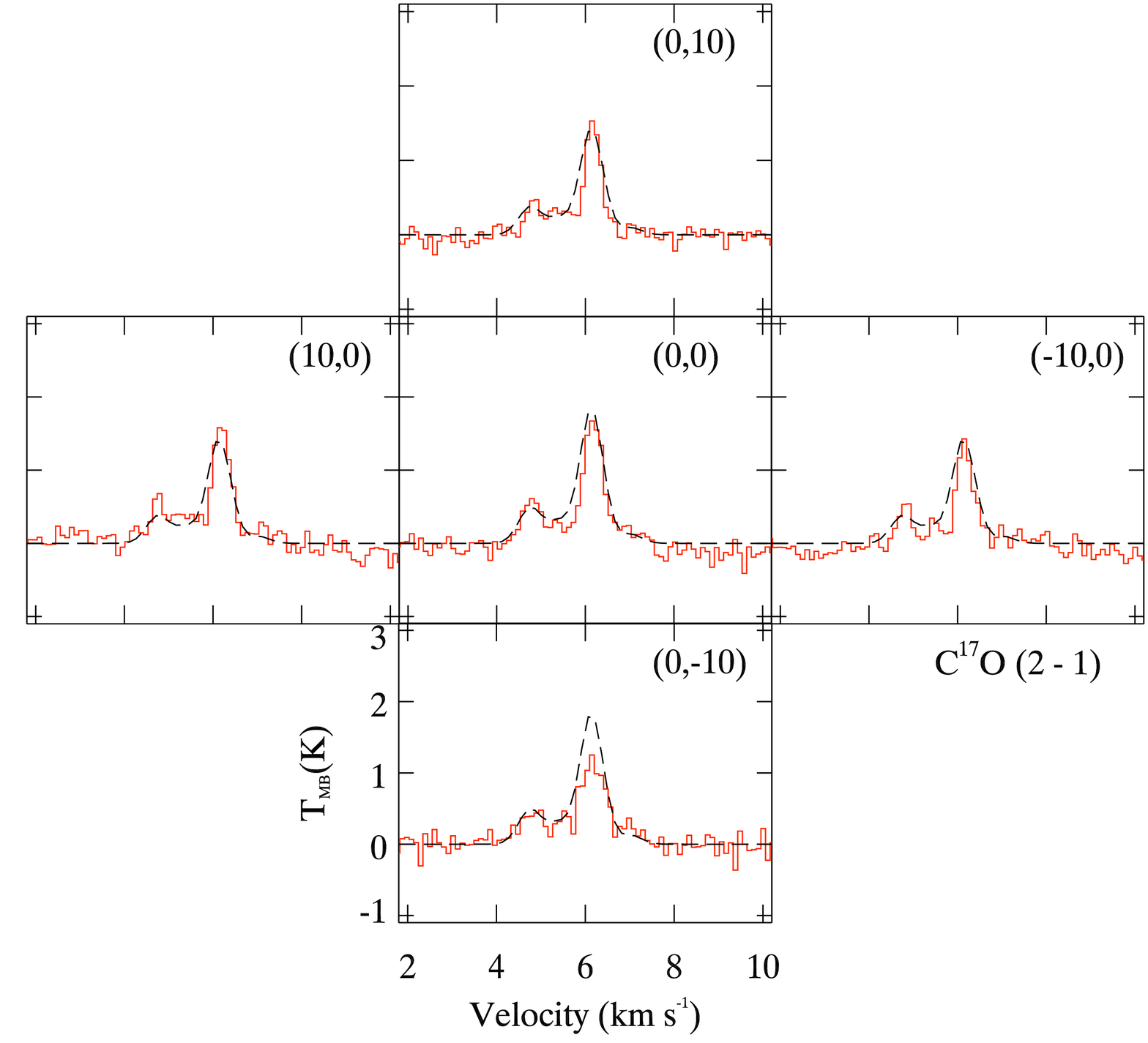}
\caption{L1527: C$^{17}$O ($J=2~$--$~1$) line profiles:
observed (solid line) and modelled (dashed line). 
The offset between cells is 10$^{\prime\prime}$.} 
\label{l1527_c17o}
\end{figure}

\begin{figure}
\centering
\includegraphics[width = 8 cm]{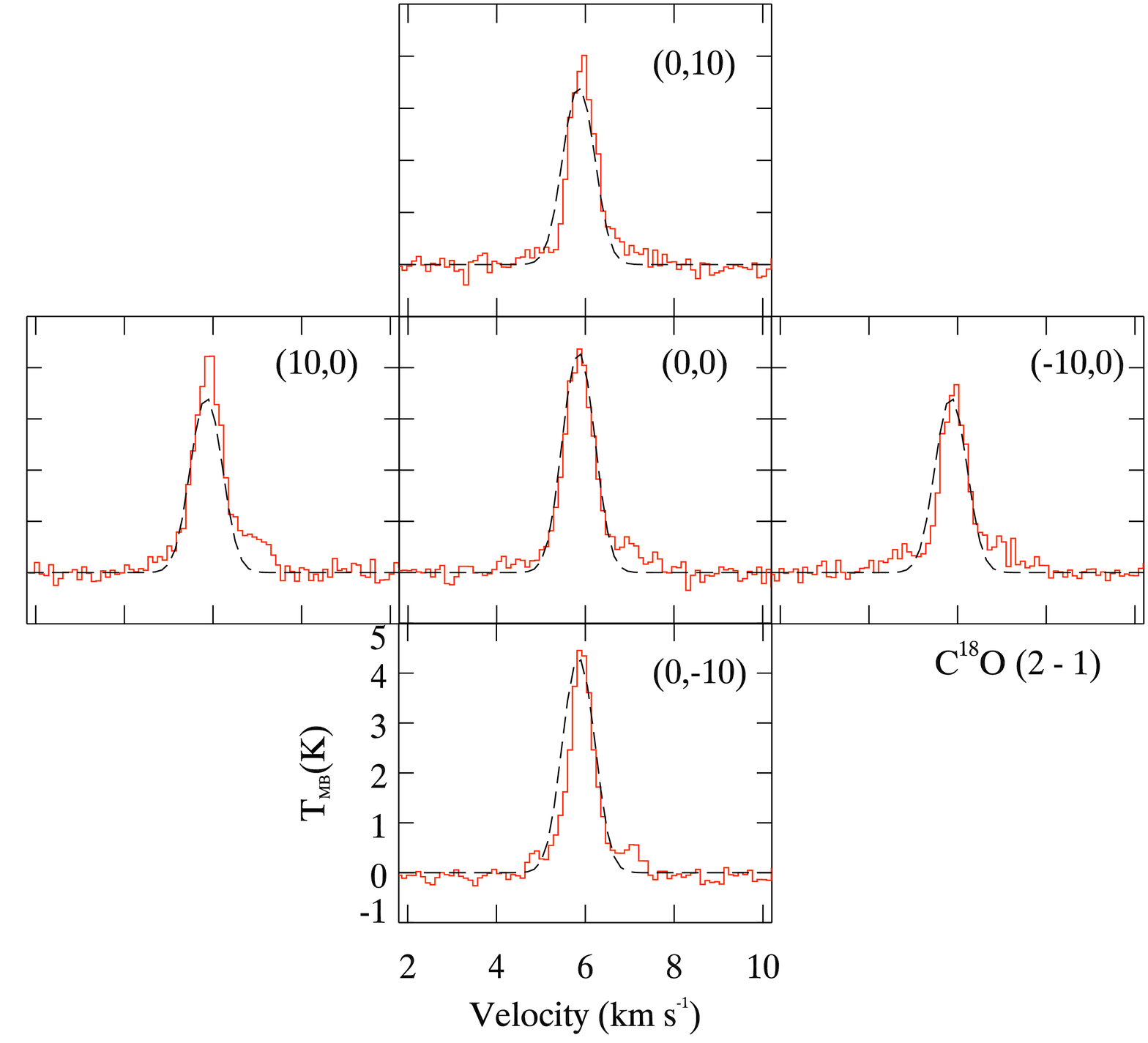}
\caption{L1527: C$^{18}$O ($J=2~$--$~1$) line profiles: 
observed (solid line) and modelled (dashed line).
The offset between cells is 10$^{\prime\prime}$.}
\label{l1527_c18o}
\end{figure}

\begin{figure}
\centering
\includegraphics[width = 8 cm]{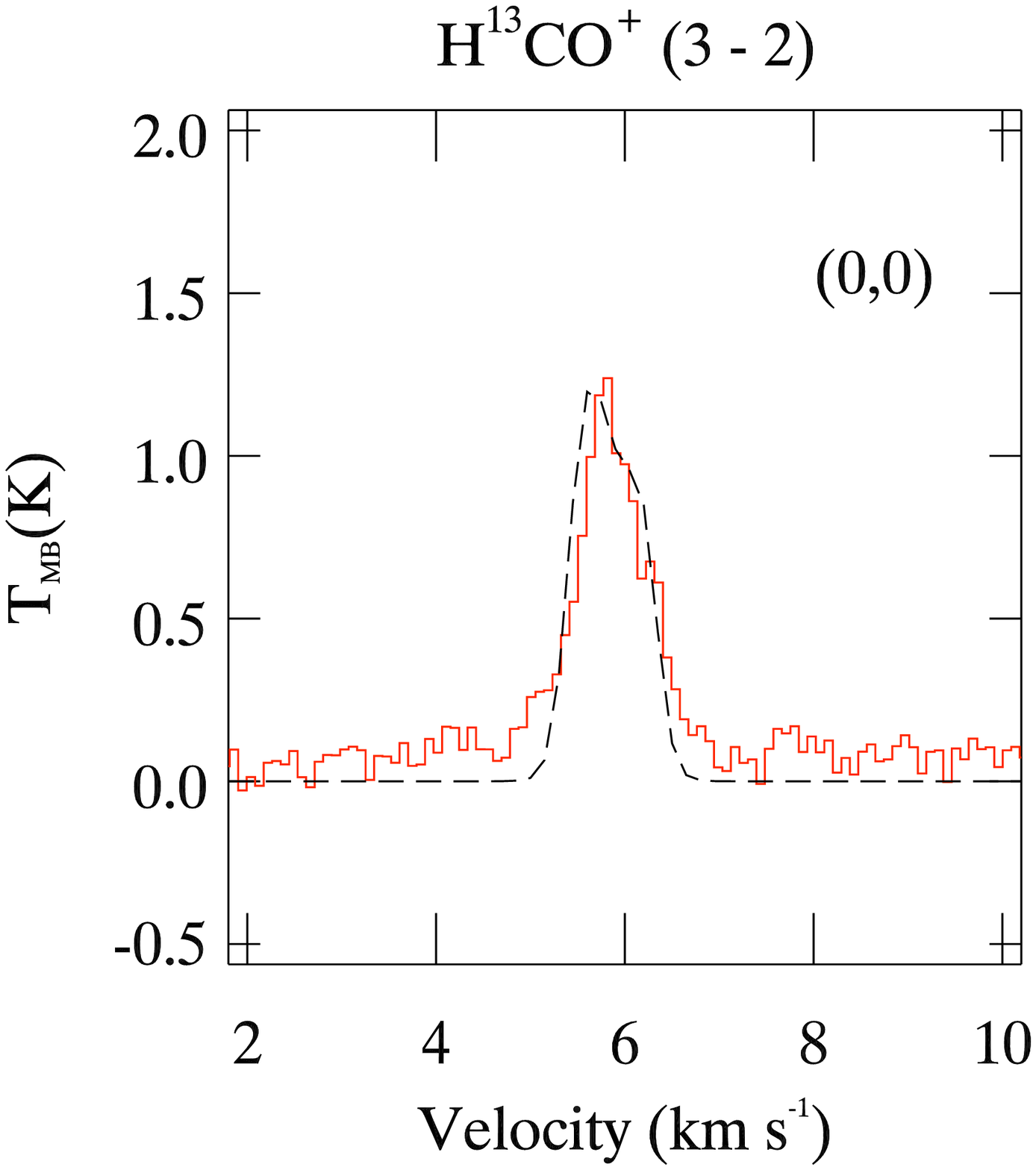}
\caption{L1527: H$^{13}$CO ($J=2~$--$~1$) line profiles:
observed (solid line) and modelled (dashed line).
The offset between cells is 10$^{\prime\prime}$.} 
\label{l1527_h13coplus}
\end{figure}

Figures~\ref{l1527_c17o} and \ref{l1527_c18o} show the modelled
and observed C$^{17}$O and C$^{18}$O ($J=2~$--$~1$) line profiles
respectively, whilst Figure~\ref{l1527_h13coplus} shows the 
H$^{13}$CO$^{+}$ ($J=2~$--$~1$) line. All three profiles are successfully
modelled as originating from the molecular gas in the envelope. Specifically,
there is no contribution to these line profiles from the gas in the molecular
outflows. The best fit parameters for these three lines are given in
Table~\ref{tab:best-fit-params-env}.

\begin{figure}
\centering
\includegraphics[width = 8 cm]{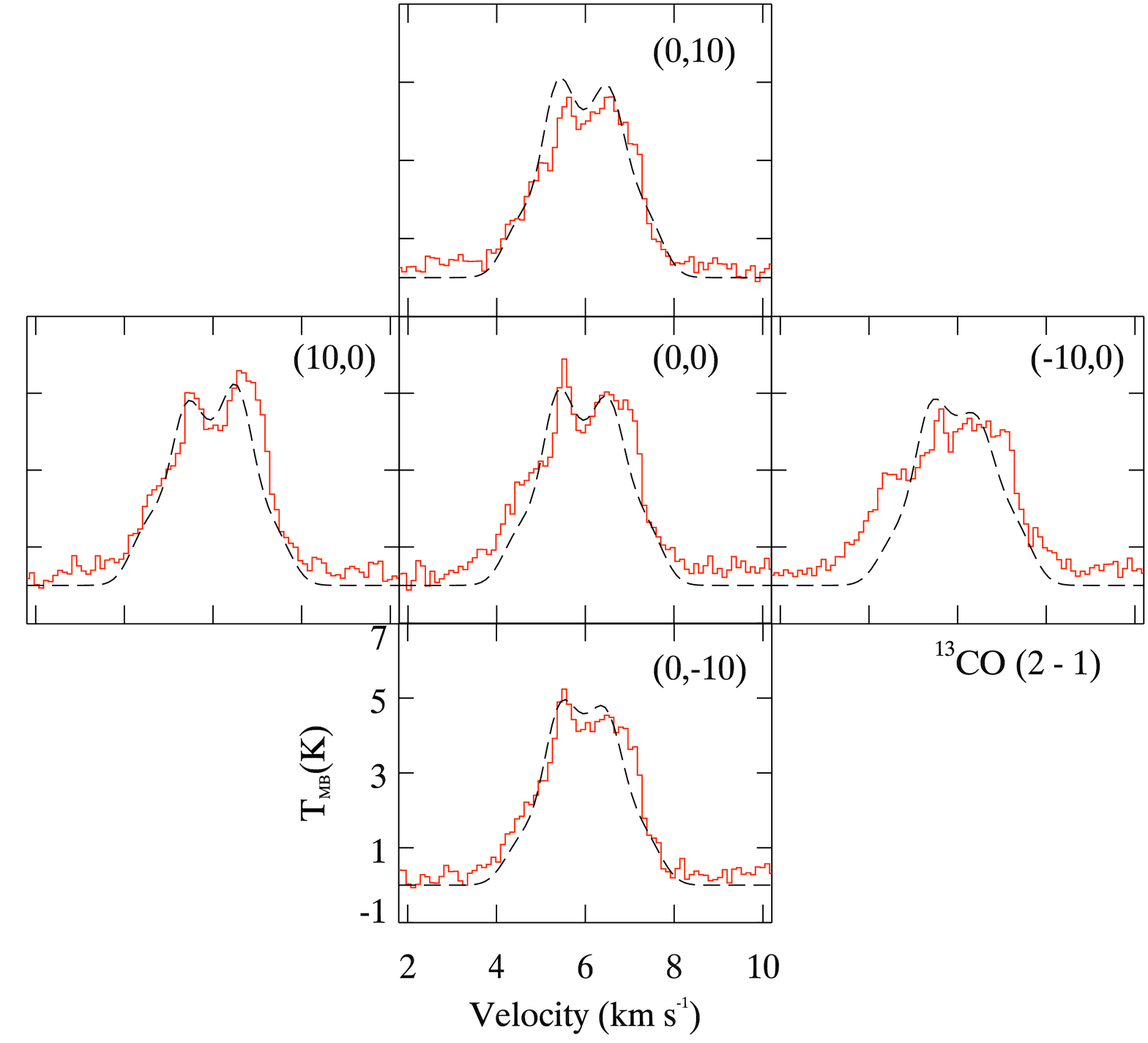}
\caption{L1527: $^{13}$CO ($J=2~$--$~1$) line profiles:
observed (solid line) and modelled (dashed line).
The offset between cells is 10$^{\prime\prime}$.} 
\label{l1527_13co}
\end{figure}

Figure~\ref{l1527_13co} shows the modelled and observed line
profiles of $^{13}$CO ($J=2~$--$~1$). The core of the line is well
fit by emission from the envelope gas. However this fails to account for 
the emission in the wings of the line profile. The best fit
parameters for the molecular gas in the outflow regions are listed
in Table~\ref{tab:best-fit-params-outer-bnd} and
\ref{tab:best-fit-params-inner-bnd}. The boundary layers
of the molecular outflow are fit with a $^{13}$CO$:^{12}$CO abundance ratio
that is significantly enhanced relative to the gas in the envelope.

\begin{figure}
\centering
\includegraphics[width = 9cm,angle=90]{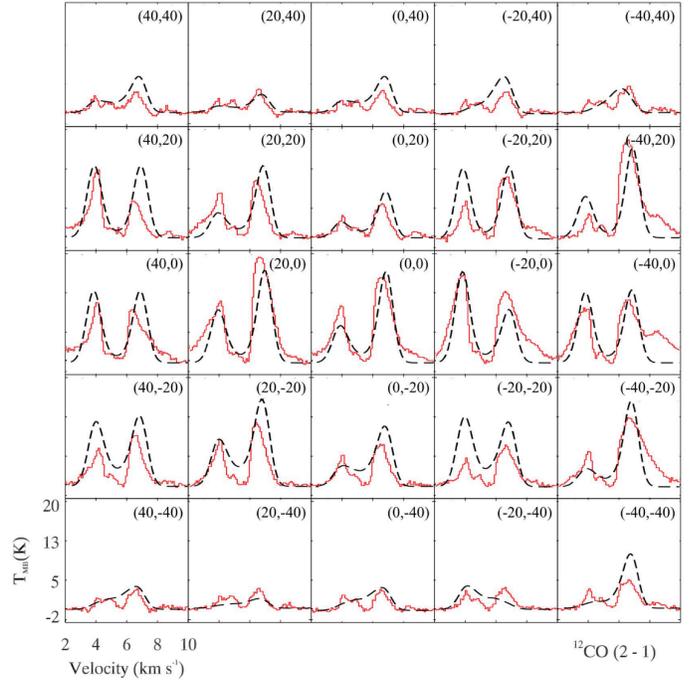}
\caption{L1527: $^{12}$CO ($J=2~$--$~1$) line profiles:
observed (solid line) and modelled (dashed line).
The offset between cells is 10$^{\prime\prime}$.}
\label{l1527_12co}
\end{figure}

Figure~\ref{l1527_12co} shows the $^{12}$CO ($J=2~$--$~1$)
line emission. This transition traces the large-scale molecular outflow, 
extending 40$^{\prime\prime}$ either side of the IRAS source which is located 
at the zero offset position. 
The line profiles are double peaked and
asymmetric with the asymmetry changing from blue to red-asymmetric
going from left to right in Figure~\ref{l1527_12co}. Again, this is
indicative of the presence of a bipolar outflow. It is consistent 
with the findings of \cite{myers.et.al95} who deduced the presence of an
outflow with a density $>$ 10$^{5}$ cm$^{-3}$ from observations of 
H$_{2}$CO (2$_{12}$ - 1$_{11}$) \& (3$_{12}$ - 2$_{11}$).

\section{Discussion}
\label{sect:discussion}

In this section the results are discussed in terms of the molecular 
gas parameters of each of the three components in our model; the 
molecular gas envelope surrounding the protostellar core and the 
inner and outer boundary layer regions of the molecular outflow. 

\subsection{Molecular Envelope}

The abundances of C$^{18}$O and C$^{17}$O are well-constrained
because the emissions originate from optically thin, quiescent, gas.
The situation is somewhat complicated by the action of freeze-out
which results in molecular depletion in the inner, denser parts of the core. 
This process is reversed at the smallest radii, due to heating by the 
protostar, but on a scale that would be undetectable at JCMT resolution.
As described above, we have quantified the level of depletion by 
comparison of dust continuum and CO line emission maps and find that 
the maximum depletion factor is $\sim$10 B335 whilst for I04166 and 
L1527 it is $\sim$5. As B335 is the densest and has the largest 
dust column density of our three sources (see 
Figure~\ref{fig:freezeout}) it is not surprising that it suffers the 
highest level of gas-phase depletion.

\subsection{Outflow abundances in inner and outer boundary layer}

The physical parameters of molecular gas in the outflow were
constrained by modelling the observed line emission of $^{12}$CO
($J=2~$--$~1$), $^{13}$CO ($J=2~$--$~1$) and HCO$^{+}$
($J=3~$--$~2$) towards each of the sources. Having large abundances,
$^{13}$CO  ($J=2~$--$~1$)  and $^{12}$CO ($J=2~$--$~1$) transitions can be 
used to trace the full extent of the outflow.
Although HCO$^{+}$ is abundant in our sources, the ($J=3~$--$~2$) 
transition has a very high critical density so that it effectively 
only traces gas deep inside the molecular envelope close to the base 
of the outflow.

Table~\ref{tab:best-fit-params-outer-bnd} and
Table~\ref{tab:best-fit-params-inner-bnd} show that the molecular 
abundances tend to be significantly larger in the boundary layers 
as compared to the cooler envelopes.
This is possibly the result of ice mantle evaporation and/or sputtering
and may also be due to enhanced gas-phase production in the warmer
environments. Thus,
\cite{rawlings.et.al04} proposed a general mechanism for molecular
enhancement of HCO$^{+}$ in a molecular outflow in which carbon atoms,
created by the photo-dissociation of newly desorbed CO gas, are 
photoionized and react with H$_2$O to form HCO$^{+}$.

Chemical fractionation apparently has a major impact on the molecular
abundance of the CO isotopes species. In \cite{carolan.et.al08} larger than expected $^{13}{\rm CO}:^{12}{\rm CO}$ ratios, compared with standard galactic values, were found for L483. This is also the case for the three sources studied here with L1527 exhibiting particularly enhanced  $^{13}{\rm CO}$ emission. The exothermic fractionation reaction leading to the creation of $^{13}$CO \citep{duley_williams} is
\begin{equation}
 \; \; \; \; \; ^{13}{\rm C}^{+} + ^{12}{\rm CO} \rightleftharpoons ^{12}{\rm C}^{+} + ^{13}{\rm CO} + \Delta{\rm E}
\label{13co_creation}
\end{equation}
where the zero-point energy difference $\Delta{\rm E}$ is equivalent to a temperature $\Delta{\rm E}/{\rm k}$ of 35~K. This mechanism is a likely source for the enhanced abundance observed from our modelling.
Alternatively, the enhanced emission from this species could be an excitation effect caused 
by the scattering of photons from the source by a moderately dense 
large extended scattering medium \citep[e.g.][]{cernicharo&guelin87}. Investigation of such an opacity effect would require further {\sc mollie} modelling of such an external scattering medium and may be appropriate for sources more deeply embedded in extended dense molecular gas.
Similar fractionation effects may also be responsible for the apparent lack of 
correlation between the abundance of C$^{18}$O and the CO depletion factor. 

The abundance of H$^{13}$CO$^{+}$ was constrained in 
B335 and L1527. The ratio of $^{13}$CO/H$^{13}$CO$^{+}$ in each
core is $\approx$ 2000 though it is slightly higher in L1527. 
The abundance of $^{13}$CO is greatly enhanced in L1527 - by a factor
of $\sim$80 (compared to a factor of $\sim$2 in B335). This indicates that the
fractionation effect discussed above for the formation of $^{13}$CO is also
feeding through to H$^{13}$CO$^{+}$, which is itself formed from $^{13}$CO.

Finally we note again that despite the wide range of line profile appearances, 
a simple outflow morphology is sufficiently robust for the purpose of 
analysing the sources. 3D modelling with a code such as {\sc mollie} enables
viewing angle effects to be isolated and reveals the underlying similarities 
and differences of the sources to be established.

\section{Summary and Conclusions}
\label{sect:conclusions}

We have used a highly simplified physical model of Class 0 protostellar 
sources, characterized by a spherically symmetric infalling envelope and
collimated bipolar outflows - coupled to a 3-D radiative transfer code - 
to model the line emission spectra from molecular gas surrounding three
sources. Physical and chemical parameters were intially constrained
from observational data and then refined by the modelling and a best fit
chi-squared analysis. 

The simplicity of the model was justified on grounds of structure 
detectability, due to the limited resolution of single dish observations.
However, using several different gas species in several transitions observed 
towards three different sources it was found that this model accurately 
and self-consistently reproduces the observed spectral line profiles.

The main finding of this study is that, although the three sources we 
investigate here have dramatically different line profile shapes, we find
that once (observationally constrained) infall, depletion and outflow components 
are introduced, a single morphological model fits all three sources.
This verifies the dominance of source morphology in determining line profile
shapes. It would be very difficult to fit the data with significantly different 
physical models, as the detailed structure of the line profiles is largely 
defined by the physics of the outflow and the envelope. The model and the 
values of its parameters are essentially defined by empirical constraints.
An alternative approach would be to formulate a more detailed theoretical 
model of the sources and then to attempt to match that with the observations.
Such an approach was adopted by \citet{dc03} in their hydrodynamic models of 
jet-driven molecular outflows. From this, they were able to identify a simple 
intensity-velocity relationship. This was successfully applied, in a 
modified exponential form, by \citet{lefloch12} to fit {\em Herschel}-HIFI 
data for five sources. Although this is a different type of modelling to 
what we have performed - and is applicable to larger scale, higher velocity 
($>10$km\,s$^{-1}$) flows - it should be possible to link consistently the 
two models.

We find that the principal cause of the source-to-source variations is
found to be the difference in the viewing angles; specifically, the angle 
between the outflow axis and the plane of the sky and {\em not} the 
assumptions about the dynamics of the inflow or outflow. 
This is evident from the large variations in the 
line profiles that we have successfully modelled as being primarily due to 
variation of the viewing angle. No variation of any other parameter or 
combination of parameters can result in such large differences.

The analysis of the multi-transistion, spatially resolved line 
profiles has yielded strong constraints on the chemical and physical 
parameters of the envelope, outflow and boundary layer of each source.
What is remarkable is that the physical parameters for each of the sources, as
presented in 
Tables~\ref{tab:best-fit-params-env}-\ref{tab:best-fit-params-inner-bnd}, are
very similar - the biggest differences between the sources being the abundance
ratios of the isotopologues.

We find strong evidence for systemic molecular depletion in all three sources.
In addition, we find that
\begin{enumerate}
\item
The C$^{18}$O$:$C$^{17}$O abundance ratio is approximately the same for 
each of the sources and is consistent with the value found in interstellar 
clouds \citep{schoier.et.al02},
\item
The abundances of $^{12}$CO and $^{13}$CO are larger in the bipolar outflow 
than in the envelope. This is probably a result of thermal evaporation 
and/or sputtering of the dust ice mantles,
\item
The ratio of the $^{13}$CO abundances in the outflow to that in the envelope
is $\sim$3 in B335 and $\sim$2 in I04166. A much larger value ($\sim$80) is 
found for for L1527. Our previous analysis for L483 yielded a value of $\sim$32.
The ratio for $^{12}$CO is $\sim$2 for both L1527 and L483. These variations 
perhaps reflect different degrees of dynamical activity in the sources, and
\item 
As found in L483 \citep{carolan.et.al08} there is evidence for $^{13}$CO 
being preferentially enhanced over $^{12}$CO in L1527, probably due to 
chemical fractionation effects.
\end{enumerate}

These studies clearly show the importance of the adoption of realistic
morphological/dynamical models of infall/outflow sources, coupled with 
full 3-D radiative transfer codes in the analysis of star-forming regions.
This approach will be essential in the analysis of ALMA data and in focussing
on the small-scale structure of the outflows. On these small scales many of our simplifying assumptions (such as that of isothermality) will no longer be applicable. In these situations, the model 
will need to be modified to include a detailed thermal structure and the 
effects of more complex morphologies and sub-structure in the outflows.

\section*{Acknowledgments}

We thank Eric Keto, Robin Garrod and David Williams for useful discussions and assistance with the observational and modelling work. We also thank an anonymous referee for constructive comments.
The James Clerk Maxwell Telescope is operated by the Joint Astronomy Centre on
behalf of the Science and Technology Facilities Council of the United Kingdom, 
the Netherlands Organisation for Scientific Research, and the National Research
Council of Canada. MPR acknowledges support from a Science Foundation Ireland Research Frontiers Programme grant (06RFP/PHY051) and from the COST Action 
CM0805 ``The Chemical Cosmos''.

\bibliographystyle{mn2e}

\appendix
\section{Estimating N(H$_2$) from C$^{18}$O observations}

The Boltzmann equation describes the level populations in local
thermodynamic equilibrium
\begin{equation}
\frac{n_{u}}{n_{l}} = \frac{g_{u}}{g_{l}} \exp
\left(-\frac{h\nu}{kT_{\rm ex}} \right),
\end{equation}
where the $n$ terms are the number density of molecules in the
given state, the $g$ terms are the statistical weights and 
$T_{\rm ex}$ is the excitation temperature. The equation of radiative
transfer can be written as
\begin{equation} T^{*}_{\rm A} = \eta_{\rm B}[J(T_{\rm ex}) -
J(T_{\rm cmb})][1 - \exp(-\tau)], \label{eqn:antenna-temp}
\end{equation}
where
\begin{equation}
J(T) = \frac{(h\nu/k)}{\exp(h\nu/kT) - 1}.
\label{eqn:rotational-transition}
\end{equation}
$T^{*}_{\rm A}$ is the antenna temperature, $\eta_{\rm B}$ is the
beam efficiency, $T_{\rm cmb}$ is the background temperature,
$\tau$ is the optical depth of the rotational transition between
the upper, $u$, and the lower, $l$, states and is given by
\begin{equation}
\tau = (n_{l}B_{lu} - n_{u}B_{ul})\frac{h\nu}{4\pi}\frac{L}{\Delta\nu}
\end{equation}
\begin{equation}
\hspace{3mm}=\frac{c^{2}}{8\pi\nu^{2}\Delta\nu}N_{l}\frac{g_{u}}{g_{l}}A_{ul}
\left[1-\exp\left(-\frac{h\nu}{kT_{\rm ex}} \right)\right].
\end{equation}
Here, $B$ and $A$ are the Einstein coefficients of absorption and
emission respectively; $L$ is the path length and so $N = nL$ is
the column density; $\Delta\nu \equiv \nu/c \Delta \upsilon$ is
the line width. Using the Boltzmann equation
\begin{equation}
\tau =
\frac{c^{2}}{8\pi\nu^{2}\Delta\nu}N_{u}A_{ul}\left[\exp\left(\frac{h\nu}
{kT_{\rm ex}}\right) - 1\right],
\end{equation}
and using the definition of $A_{ul}$
\begin{equation}
A_{ul} = \frac{64\pi^{4}\nu^{3}}{3hc^{3}}\frac{\mu^{2}S}{g_{u}},
\end{equation}
where $\mu$ is the permanent dipole moment and S is the line
strength gives
\begin{equation}
\frac{N_{u}}{g_{u}} =
\frac{3h}{8\pi^{3}}\frac{1}{\mu^{2}S}\frac{\tau\Delta\upsilon}
{[\exp(h\nu/kT_{\rm ex})-1]}.
\end{equation}
The column density in the transition is related to the total
column density of the species by
\begin{equation}
\frac{N_{u}}{g_{u}} = \frac{N_{\rm tot}}{Q(T_{\rm rot})}
\exp\left(-\frac{E_{u}}{T_{\rm rot}} \right)
\end{equation}
where Q($T_{\rm rot}$) is the partition function, E$_{\rm u}$ is the energy
of the upper level and $T_{\rm rot}$
is the rotational temperature, assumed to be the same for all
levels. It is usually further assumed that $T_{\rm rot} = T_{\rm ex}$ so that
\begin{equation}
N_{\rm tot} = \frac{3h}{8\pi^3}\frac{Q(T_{\rm ex})}{\mu^{2}S}
\tau\Delta\upsilon\frac{\exp(E_{u}/kT_{\rm ex})}{\exp(h\nu/kT_{\rm ex})-1}. 
\label{eqn:col-den-derivation}
\end{equation}
This expression can be used if $\tau$, $\Delta\upsilon$ and $T_{\rm ex}$
are known.

Alternatively, if the optical depth is small then
Equation~(\ref{eqn:antenna-temp}) becomes
\begin{equation}
T^{*}_{\rm A}/\eta_{\rm B} \simeq [J(T_{\rm ex}) - J(T_{\rm cmb})]\tau
\end{equation}
and substituting for $\tau$ in
Equation~(\ref{eqn:col-den-derivation}) and using
Equation~(\ref{eqn:rotational-transition}) gives 
\begin{eqnarray}
N_{\rm tot} &=& \frac{3k}{8\pi^{3}\nu}\frac{Q(T_{\rm ex})}{\mu^2S}
\frac{J(T_{\rm ex})}{J(T_{\rm ex})-J(T_{\rm cmb})}
\nonumber \\
&\times& \exp \left(\frac{E_u}{kT_{\rm ex}} \right) \int T_{\rm mb} d\upsilon
\end{eqnarray} 
where we have also used 
$T^{*}_{\rm A}\Delta\upsilon/\eta_{\rm B}\simeq \int T_{\rm mb} d\upsilon$,
the integrated line intensity.
Finally, if $T_{\rm ex} >> T_{\rm cmb}$ this can simplified further to
\begin{equation}
N_{\rm tot} = \frac{3k}{8\pi^{3}\nu}\frac{Q(T_{\rm ex})}{\mu^2S}\exp 
\left(\frac{E_u}{kT_{\rm ex}} \right) \int T_{\rm mb} d\upsilon.
\end{equation}
If the integrated line strength is measured in km s$^{-1}$,
frequency in GHz, $\mu$ in debye 
then the column density, in CGS units, of an optically thin gas with 
negligible contribution from the cosmic microwave background is
\begin{eqnarray}
N_{\rm tot} &=& 1.67 \times 10^{14}~{\rm cm^{-2}}~\frac{Q(T_{\rm ex})}
{\nu\mu^{2}S} \nonumber \\
&\times& \exp \left(\frac{E_u}{kT_{\rm ex}} \right) \int T_{\rm mb} d\upsilon.
\end{eqnarray}

In our models an excitation temperature of 10K is assumed (consistent
with the dust temperature) yielding a C$^{18}$O rotational partition 
function of $\sim 7.94$.
$\mu$ is the dipole moment which for C$^{18}$O is 0.11079 Debye
\citep{pickett.et.al98}. A conversion factor of
2.07~$\times$~10$^{6}$ is used to convert the C$^{18}$O column
density into the H$_{2}$ column density
\citep{ladd.et.al98,lacy.et.al94,wilson&rood94}.

\end{document}